\documentclass{IEEEtran}
\usepackage{cite}
\usepackage{amsmath,amssymb,amsfonts}
\usepackage{algorithmic}
\usepackage{graphicx}
\usepackage{color}
\usepackage{pdfpages}
\usepackage{bm}
\usepackage{array}
\usepackage{url}

\ifCLASSOPTIONcompsoc
    \usepackage[caption=false,font=footnotesize,labelfont=rm,textfont=rm]{subfig}
\else
\usepackage[caption=false, font=footnotesize]{subfig}
\fi
\usepackage{textcomp}

\def\BibTeX{{\rm B\kern-.05em{\sc i\kern-.025em b}\kern-.08em
    T\kern-.1667em\lower.7ex\hbox{E}\kern-.125emX}}

\newcommand{\iz}{\hat{\bm\imath}_z}
\newcommand{\irho}{\hat{\bm\imath}_\rho}
\newcommand{\iphi}{\hat{\bm\imath}_\phi}

\newcommand{\normal}{\hat{\bm\imath}_n}

\begin{document}
\title{A Body-of-Revolution Human Model for RF Sensing with Measurement-Driven Calibration for Indoor Environments}
\author{Haoqing Wen, Michele D'Amico, \IEEEmembership{Senior Member, IEEE}, Matteo Oldoni, \IEEEmembership{Member, IEEE}, Federica Fieramosca,  \IEEEmembership{Graduate Student Member, IEEE}, Vittorio Rampa, \IEEEmembership{Senior Member, IEEE}, Stefano Savazzi, \IEEEmembership{Senior Member, IEEE}, Qi Wu, \IEEEmembership{Member, IEEE} and Gian Guido Gentili, \IEEEmembership{Member, IEEE}
\thanks{This work was supported in part by the China Scholarship Council program (Project ID: 202506020097), the National Science Foundation of China under grant 62525104, and was supported in part by the European Union under the Italian National Recovery and Resilience Plan (PNRR) of NextGeneration EU, partnership on “Telecommunications of the Future” (PE00000001 - program “RESTART”, Structural Project SRE). (\textit{Corresponding
author: Gian Guido Gentili.})}
\thanks{Haoqing Wen and Qi Wu are with the School of Electronics and Information Engineering,
Beihang University, Beijing 100083, China, and also with the Zhongguancun
Laboratory, Beijing 100094, China (e-mail: whq001110@buaa.edu.cn; qwu@buaa.edu.cn).}
\thanks{Gian Guido Gentili, Michele D’Amico, Matteo Oldoni, and H. Wen are with the Dipartimento di Elettronica, Informazione e Bioingegneria (DEIB), Politecnico di Milano, 20133 Milan, Italy (e-mail: gianguido.gentili@polimi.it; michele.damico@polimi.it; matteo.oldoni@polimi.it).}
\thanks{Stefano Savazzi, Vittorio Rampa, and Federica Fieramosca are with the Institute of Electronics, Computer and Telecommunication Engineering (IEIIT), National Research Council of Italy (CNR), 20133 Milan, Italy (e-mail: stefano.savazzi@cnr.it; vittorio.rampa@cnr.it; federica.fieramosca@cnr.it).}}

\maketitle

\begin{abstract}
Model training for Device-Free Localization (DFL) and Radio-Frequency (RF) sensing systems heavily relies on large-scale datasets, which are costly and time-consuming to obtain through measurements across different environments and sensing configurations. Lightweight yet physically consistent propagation models are therefore critical for efficient generation of realistic RF sensing data. This paper presents an RF sensing prediction approach for indoor environments based on a Body of Revolution (BoR) human model. A fast 2.5-Dimensional Finite Element Method (2.5-D FEM) is proposed for computing the scattering fields of a human-like BoR model under the excitation of a vertical polarized dipole. Through comparisons, the proposed BoR model is shown to preserve scattering characteristics close to 3-D human bodies while yielding a smaller computational cost compared to a simple cylindrical model. A measurement-driven background-field modeling approach is further introduced for practical indoor applications, accounting for the complex propagation effects of indoor environments implicitly. Comparing with measurements of a typical indoor DFL scenario, the proposed approach achieves approximately 85\% prediction accuracy and reproduces the spatial Received Signal Strength Indicator (RSSI) variations observed in practice, proving its potential for RF sensing prediction and large-scale database generation at a fraction of the computational cost required for full-wave simulations.
\end{abstract}


\section{Introduction}

\begin{figure}[!t]
\centerline{\includegraphics[width=\columnwidth]{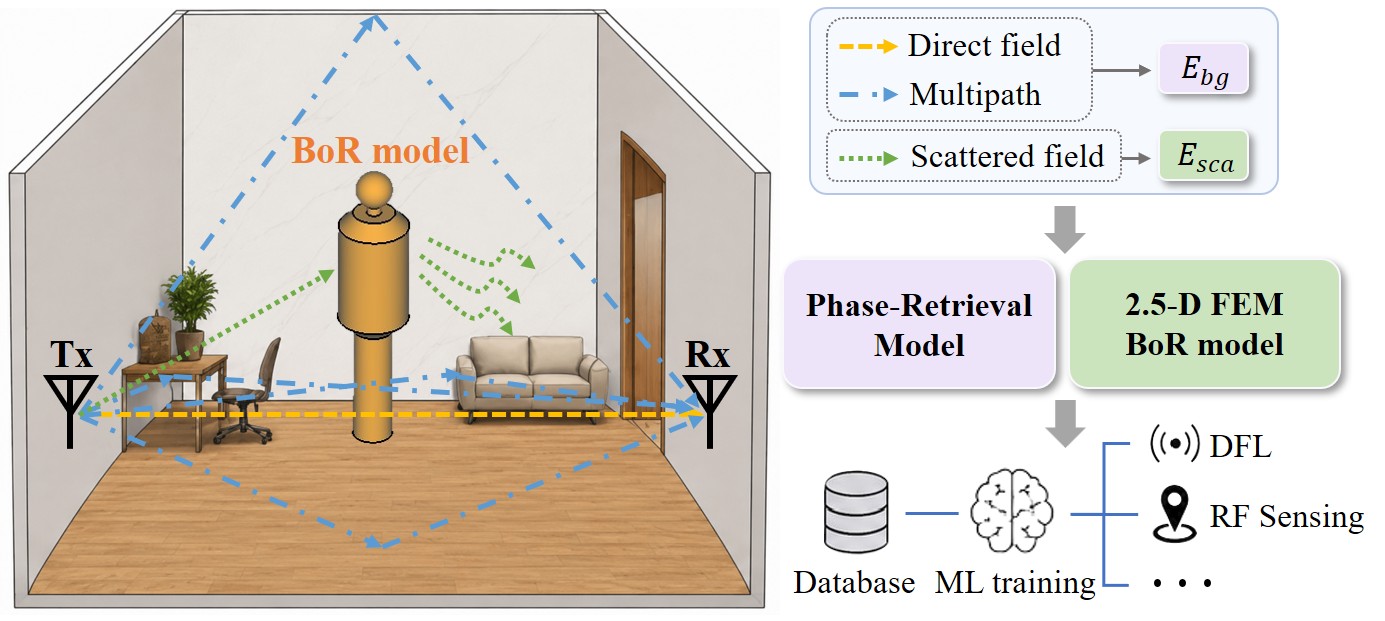}}
\caption{Overview of the proposed RF sensing prediction framework. Human-induced scattering fields are computed using a 2.5-D FEM solver for a BoR human model, while the background field is reconstructed via a measurement-driven phase-retrieval approach. Their superposition yields RSSI predictions for RF sensing analysis and large-scale database generation.}
\label{fig.abstract}
\end{figure}

\IEEEPARstart{R}{adio}-Frequency (RF) sensing and Device Free Localization (DFL) have attracted significant attention in recent years due to their potential in a wide range of indoor applications, including smart environments~\cite{DFL_smartcity}, multi-target recognition~\cite{tomography,localization}, health care~\cite{DFL_healthcare,healthcare02}, and human activity recognition~\cite{DFL_HumanActive}. Unlike conventional wireless systems that rely on active devices carried by users, DFL techniques, also known as passive localization methods, achieve target sensing by exploiting the perturbations induced by human presence or motion on the surrounding electromagnetic (EM) field distributions and scattering characteristics~\cite{Fieramosca_awpl}, or on the wireless propagation channel~\cite{Stefano2019}. The human body acts as a complex scatterer whose EM properties and spatial variations lead to measurable changes in the Received Signal Strength Indicator (RSSI)~\cite{RSSI_meas}, Channel State Information (CSI)~\cite{CSI_attention,CSI_CNN}, or other RF observables~\cite{review2017}. Accurate characterization of human-induced scattering effects is therefore a fundamental problem in the design and analysis of RF sensing systems.

With the widespread adoption of data-driven and Machine Learning (ML) techniques in RF localization and sensing, a variety of methods have been proposed, such as Bayesian inference~\cite{ML_Bayesian}, Variational AutoEncoders (VAE)~\cite{ML_VAE_Stefano,ML_VAE_Federica}, and Generative Neural Network (GNN)~\cite{ML_GNN_Stefano,ML_GNN_Barba,ML_review}, to detect and track human positions, movements, and activities. Most of these methods rely on approximate knowledge of a physical (prior) model to interpret the perturbations in radio signals induced by human presence or motion. Although tremendous progress has been achieved in ML-based RF sensing, the availability of large-scale, diverse, and well-labeled datasets remains one of the major challenges limiting their widespread adoption \cite{ML_RFSensing_Review}. Ideally, training data should cover a broad range of human positions, deployment configurations, and environmental conditions to ensure adequate model generalization. However, collecting such datasets requires extensive measurement campaigns involving multiple subjects, different RF device setups, and various indoor environments, thus resulting in substantial costs in terms of time and labor spent during data collection. The automation of data measurement campaigns is another key factor that hinders the widespread adoption of DFL techniques in several scenarios. Therefore, developing physically consistent approaches that efficiently generate RF sensing data has become an important research direction to reduce reliance on large-scale measurements.

One promising approach to alleviate the reliance on extensive measurements is the use of EM modeling and numerical simulation techniques. By predicting the interaction between human bodies and wireless signals through physics-based models, synthetic RF sensing datasets can be generated under a wide range of scenarios, thereby complementing or partially replacing costly measurement campaigns. Existing approaches can be broadly classified into three categories. Full-wave EM methods, such as the three-dimensional (3-D) Finite Element Method (FEM)~\cite{3DFEM}, Finite-Difference Time-Domain (FDTD)~\cite{FDTD01,FDTD02}, and Method of Moments (MoM)~\cite{MOM01,MOM02}, provide high modeling accuracy but require extraordinary computational time and resources, which limits their applicability to large-scale database generation. Analytical or semi-analytical models approximate the human body using canonical geometries, such as spheres \cite{GreenHead}, infinite cylinders \cite{AnaCylind01}, finite-length cylinders \cite{AnaCylind03}, and elliptical cylinders \cite{AnaCylind02,Plouhinec2023Cylinder}, offering high efficiency at the cost of severely limited realism in representing human scattering characteristics. High-frequency approximation methods \cite{hf_method01,hf_method02,Vittorio202201,Vittorio202202} offer computational efficiency for large-scale propagation analysis, but may lack accuracy in modeling near-field interactions and human-body scattering. Consequently, existing simulation methods still face a trade-off between efficiency and accuracy, making them inadequate for generating large-scale, high-fidelity datasets required by machine-learning-based RF systems.

The Body of Revolution (BoR) human model provides a compromise between simplified cylindrical approximations and computationally intensive 3-D human models. Preserving the major geometric features of the human body, it enables the use of an efficient 2.5-dimensional Finite Element Method (2.5-D FEM) for full-wave EM analysis. Although 2.5-D FEM has been extensively studied for BoR structures~\cite{Jinjianming_borfem01,Jinjianming_borfem02,BOR_FEM,2.5DFEM02,2.5DFEM03,2.5DFEM04}, most existing formulations are developed for plane-wave excitation. More importantly, the applicability of BoR human models has not been systematically validated in RF sensing scenarios.

The modeling of the indoor propagation environment also plays a crucial role in determining the RSSI in RF sensing applications. In practical scenarios, wireless signals are affected by multipath propagation as well as reflections from walls, floors, ceilings, and surrounding objects, resulting in complex background fields that are difficult and often impossible to characterize accurately. Empirical statistical models and deterministic EM methods are two categories of common indoor channel modeling approaches. The former are computationally efficient but often lack sufficient accuracy for a specific environment \cite{statistical_channel}. The latter, such as ray-tracing methods \cite{RT_revirew,RT01}, can achieve high prediction accuracy but rely heavily on detailed knowledge of the environment geometry and material properties, and involve considerable computational complexity for repeated field predictions.

Unlike previous approaches, this paper presents a measurement-driven modeling approach for indoor RF sensing environments, in which the background field is reconstructed from a small sample of measurements. By superimposing the reconstructed background field and the scattering field of a BoR human model computed by the proposed 2.5-D FEM, fast and accurate prediction of RSSI variations in indoor RF sensing scenarios can be achieved.

Specifically, the contributions of this work are threefold: 
\begin{itemize}
    \item \textit{Efficient 2.5-D FEM for dipole-excited BoR human model:} a fast 2.5-D FEM formulation for dipole-excited human-like BoR structures is proposed to compute body scattering. The numerical accuracy of the proposed approach is validated through comprehensive comparisons with the commercial EM simulator Feko$^\text{TM}$ \cite{FEKO2025}.
    \item \textit{Assessment of the BoR human model for RF sensing simulations:} through comparisons with both cylindrical and 3-D human models, the proposed BoR human model is verified to retain the principal scattering characteristics at a lower computational cost. Furthermore, the applicability of this model in a practical RF sensing scenario is validated by the good agreement with measurements.
    \item \textit{Background-field reconstruction via phase retrieval:} a novel approach is proposed to reconstruct the unknown background-field phases in indoor RF sensing environments from a small sample of measurements. The effects of indoor multipath propagation and environmental reflections are implicitly encoded in the complex background field using background measurements with a small set of human-presence measurements in the same environment.
\end{itemize}

The rest of the paper is organized as follows. Section~\ref{chap:method} presents a 2.5-D FEM for EM scattering computation of a BoR human model under a vertical polarized dipole excitation. In Section~\ref{chap:validation}, we validate the proposed 2.5-D FEM through comparisons with the Feko$^\text{TM}$ simulations and investigate the effectiveness of the BoR human model in balancing scattering fidelity and computational efficiency. The indoor DFL scenario is introduced in Section~\ref{chap:experimental_results}, where we present the measurement-driven environment modeling approach, and validate the proposed framework through comparisons with measurements. Finally, Section~\ref{chap:conclusion} draws some conclusions.

\section{Numerical Method}
\label{chap:method}

The EM problem considered here involves an arbitrarily shaped BoR illuminated by a vertically oriented Hertzian dipole source. The BoR approximation allows decoupling the contributions of different harmonics in which the field can be decomposed, thereby significantly reducing the computational time required to solve the associated linear system of equations derived from the FEM application. Perfectly Matched Layer (PML) \cite{berenger2007perfectly} is used in proximity of the BoR to accurately simulate free-space, so that the domain of FEM computation is limited to a rather small region around the BoR. In order to compute the field at any distance from the BoR, we deploy equivalent sources all around the body and use them for field computation outside of the FEM domain. This also contributes to the significant speed of analysis of the overall methodology.

\subsection{2.5-D FEM}

The BoR is centered in a local cylindrical reference frame $(\rho,\phi,z)$, with unit vectors $\irho$, $\iphi$, $\iz$. The source that provides the forcing term is a $z$-directed Hertzian dipole placed outside the FEM domain and, with no loss of generality, on the $\phi=0$ half plane. Within the scattered field formulation of the 2.5-D FEM problem, the scattered EM field is expanded as a sum of circular harmonics in $\phi$ and solved in the $\rho, z$ plane. In that plane, we label ${\bf E}_t$ the in-plane component of the electric field ($\rho$-$z$ components) and ${\bf E}_\phi$ the out-of-plane component. Using the superscript $(m)$ to refer to the generic harmonic in $\phi$, with $M$ the number of harmonics, we can write the following expansion for the scattered electric field ${\bf E}$ as
\begin{equation}\label{eq:Et_Ephi}
{\bf E}(\rho,\phi,z)
=
\sum_{m=0}^M{\bf e}_t^{(m)}(\rho,z)c_m+\iphi\sum_{m=1}^M {e}_\phi^{(m)}(\rho,z)s_m,
\end{equation}
where 
\begin{equation}
c_m = \frac{1}{\sqrt{\pi\varepsilon_{m}}}\cos m\phi,
\end{equation}
\begin{equation}
s_m = \frac{1}{\sqrt{\pi}}\sin m\phi,
\end{equation}
\begin{equation}
\varepsilon_{m}=\left\{
\begin{array}{ll}
2 & \text{if } m=0 \\
1 & \text{if } m> 0
\end{array}.
\right.
\end{equation}

The weak form of the scattered field formulation of FEM \cite{Jinbook} in the domain $V$ for penetrable scatterers can be written as 
\begin{align}\label{eq:weak_form}
        \int_V \nabla\times{\bf W}\cdot\nabla\times{\bf E}\; dV - k_0^2\int_V \epsilon_r {\bf W}\cdot{\bf E}\;dV \nonumber \\
         =k_0^2\int_{V} (\epsilon_r - 1){\bf W}\cdot{\bf E}_i \;dV,
\end{align}
where ${\bf E}$ is the scattered electric field, ${\bf H}$ is the corresponding magnetic field, ${\bf W}$ is an arbitrary testing function belonging to $H(\text{curl})$ and satisfying essential boundary conditions, and ${\bf E}_i$ is the incident field generated by a source placed outside the FEM domain. 
The detailed derivations of the weak-form discretization procedure by Galerkin's method for \eqref{eq:weak_form} are provided in Appendix \ref{app:fem}. Following the FEM discretization, a sparse block diagonal matrix problem is found, in which each mode corresponding to a value $m$ of the azimuthal index is solved separately as 
\begin{equation}\label{eq:main_m}
    \mathbb{U}^{(m)} = k_0^2\left(\mathbb{A}^{(m)}-k_0^2\mathbb{B}' \right)^{-1}\mathbb{K}^{(m)}.
\end{equation}
Details of the matrix expression and the block diagonal structure are shown in the Appendix.

\subsection{Known term: $z$-directed Hertzian dipole}

The method can handle a quite arbitrary source, but, in order to match the practical transmitting and receiving antennas in typical DFL scenarios, an arbitrarily placed $z$-directed Hertzian dipole was chosen as the excitation source. Such a source provides the incident electric field ${\bf E}_i$ of the scattered field formulation of FEM in \eqref{eq:weak_form}.  Without loss of generality, the dipole is assumed to be located at ${\bf r}_s =(\rho_s,0,z_s)$, with a source moment ${\bf p} = \iz Il$. From the knowledge of the power $P_r$ radiated by the source, we can obtain the source moment component (assumed real) as
\begin{equation}
    Il = \sqrt{\frac{3P_r\lambda_0^2}{\pi\eta_0}},
\end{equation}
where $\lambda_0$ is the  free-space wavelength and $\eta_0 = 376.73$ $\Omega$ is the free-space impedance. Assuming that the dipole is not very close to the human body (i.e, $R = |{\bf r}-{\bf r}_s| > 3\lambda_0$), we can write the incident field at ${\bf r}$ as 
\begin{equation}\label{eq:zdipole_close}
\mathbf{E}_i(\mathbf{r})
= -j\omega\mu_0 
 \frac{e^{-jkR}}{4\pi R}\,
\mathbf{p}_\perp.
\end{equation}
The subscript '$\perp$' is a compact form for
\begin{equation}\label{eq:perp1}
    {\bf p}_\perp = \frac{{\bf r}-{\bf r}_s}{R}\times{\bf p}\times\frac{{\bf r}-{\bf r}_s}{R}.
\end{equation}

The field expressions thus obtained are now decomposed into azimuthal Fourier modes in $\phi$. We can write 
\begin{equation}
    {\bf E}_i = \sum_{m=0}^M {\bf e}_{i,t}^{(m)}c_m + \iphi\!\sum_{m=1}^M {e}_{i,\phi}^{(m)}s_m,
\end{equation}
where ${\bf e}_{i,t}^{(m)} = \irho e_{i,\rho}^{(m)} + \iz e_{i,z}^{(m)}$, to find
\begin{align}
    & e_{i,\rho}^{(m)}(\rho,z)=  \int_0^{2\pi} \irho\cdot{\bf E}_i({\bf r}) c_m d\phi,\label{eq:Er_m}\\
    & e_{i,z}^{(m)}(\rho,z)=  \int_0^{2\pi} \iz\cdot{\bf E}_i({\bf r}) c_m d\phi,\label{eq:Ez_m}\\
    & e_{i,\phi}^{(m)}(\rho,z)=\int_0^{2\pi}\iphi\cdot{\bf E}_i({\bf r})s_m d\phi.\label{eq:Ep_m}
\end{align}
From the decomposition in Fourier harmonics, we can obtain the expression of the known term in \eqref{eq:main_m}. We have
\begin{equation}
    \mathbb{K}_t^{(m)}(p) = \int_S (\epsilon_r - 1) {\bm\tau}_p\cdot {\bf e}_{i,t}^{(m)}\rho d\rho dz,
\label{eq:tau}
\end{equation}
\begin{equation}
    \mathbb{K}_\phi^{(m)}(p) = \int_S (\epsilon_r - 1) {\varphi}_p { e}_{i,\phi}^{(m)}d\rho dz.
\label{eq:phi}
\end{equation}
In \eqref{eq:tau}-\eqref{eq:phi}, ${\bm\tau}_p$ is the generic 2-D edge elements and $\varphi_p$ is the generic Lagrange 2-D nodal element.
The interested reader may refer to Appendix \ref{app:fem} for the complete description of the BoR-FEM and the symbols employed in this section.

\subsection{External field computation by equivalent sources}

The solution of \eqref{eq:main_m} yields the scattered electric and magnetic  fields everywhere in the FEM domain as
\begin{equation}\label{eq:fieldFEME}
    {\bf E} = \sum_{m=0}^M\sum_{q=1}^Q u_{q}^{(m)}{\bm\tau}_qc_m + \iphi\sum_{m=1}^M\sum_{q=1}^{Q'} v_{q}^{(m)}\frac{\varphi_q}{\rho} s_m,
\end{equation}
\begin{align}\label{eq:fieldFEMH}
    {\bf H} =& \left\{\sum_{m=0}^M\sum_{q=1}^Q u_{q}^{(m)}\left[\iphi(\nabla\times{\bm\tau}_q)_\phi c_m-\frac{m}{\rho} {(\nabla\times{\bm\tau}_q)_t}s_m\right]  \nonumber \right. \\ 
    & +\left. \sum_{m=1}^M\sum_{q=1}^{Q'} v_{q}^{(m)}{(\nabla\times\iphi\varphi_q)_t} s_m \right\}\frac{1}{-j\omega\mu_0},
\end{align}
where, to compact the notation, we set
\begin{equation}
    (\nabla\times\bm\tau)_t = {\tau_x\irho -\tau_\rho\iz},
\end{equation}
\begin{equation}
    (\nabla\times\bm\tau)_\phi =\frac{\partial\tau_\rho}{\partial z}-\frac{\partial \tau_z}{\partial \rho},
\end{equation}
\begin{equation}\label{eq:nablavarphi}
    (\nabla\times\iphi\varphi)_t = 
    \frac{\partial\varphi}{\partial\rho}\iz - \frac{\partial\varphi}{\partial z}\irho.
\end{equation}

Starting from (\ref{eq:fieldFEME})--(\ref{eq:fieldFEMH}), the scattered fields at all points outside the FEM domain can be computed by using equivalent sources. This post-processing solves the issue of discretizing by FEM a large space around the human body to compute the field at all receivers in the environment, which can be placed at a significant distance from the human body. A cylindrical surface $S_\text{eq}$ in proximity of the human body is defined, and the scattered fields on this surface are computed by (\ref{eq:fieldFEME})--(\ref{eq:fieldFEMH}). An outward normal unit vector $\normal$ is introduced on $S_\text{eq}$ and surface equivalent electric (${\bf J}_s$) and magnetic (${\bf M}_s$) current distributions are deployed on $S_\text{eq}$. They are related to the fields by 
\begin{equation}\label{eq:JM}
    {\bf J}_s = \normal\times{\bf H},
\end{equation}
\begin{equation}\label{eq:M}
     {\bf M}_s = -\normal\times{\bf E}.
\end{equation}
The electric field at all points outside $S_\text{eq}$ is then computed by
\begin{align}\label{eq:eqsources}
    {\bf E}({\bf r}) = -j\omega\mu\int_{S_\text{eq}}{\bf J}_{s\perp}\frac{e^{-jkR'}}{4\pi R'}dS'  \nonumber \\  + jk\int_{S_\text{eq}}\frac{{\bf R}'}{R'}\times{\bf M}_{s}\frac{e^{-jkR'}}{4\pi R'} dS',
\end{align}
where subscript '$\perp$' was defined in \eqref{eq:perp1}, ${\bf R}' = {\bf r}-{\bf r}'$, and $R'=|{\bf R}'|$, being ${\bf r}'$ the equivalent source position. 

It should be emphasized that the equivalent source formula in \eqref{eq:eqsources} is not used as an approximation for the entire computational domain. Instead, the FEM region is defined as a finite domain surrounding the BoR and extending several wavelengths away from it, within which the EM field is rigorously computed using full-wave analysis. The equivalent sources are used only to reconstruct the field outside the FEM domain, at distances of several wavelengths from the equivalent surface. The proposed hybrid strategy can accurately evaluate the entire spatial region while maintaining high computational efficiency. The numerical accuracy of the proposed method will be strictly validated in the next section.

\section{Validation of 2.5-D FEM and Assessment of BoR Human Model}
\label{chap:validation}

We discuss first in this section a numerical verification of the proposed 2.5-D FEM for scattering computation, followed by the advantages of the BoR human model in balancing informativeness and efficiency. Rather than introducing specific application scenarios, this section aims to establish the general validity and significance of approximating the human body as a BoR model for EM scattering analysis.

\subsection{Validation of 2.5-D FEM}
\label{subsec:validation}

\begin{figure}[!t]
\centerline{\includegraphics[width=3.3in]{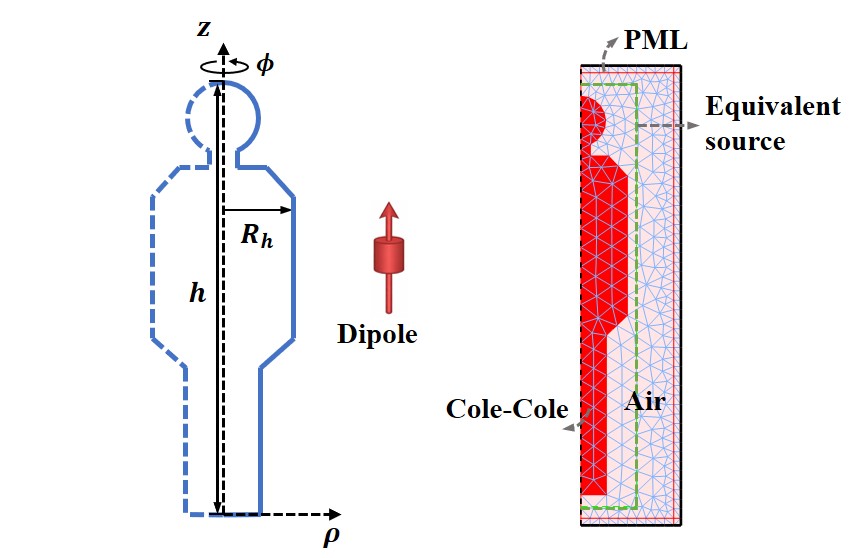}}
\caption{Schematic of the human body under the excitation of a $z$-directed  dipole, including an illustrative 2.5-D FEM mesh with labeled material regions (the mesh is shown for illustration only, a finer mesh is employed in the computations).}
\label{fig.body}
\end{figure}

In the following discussion, we used the human body model shown in Fig.~\ref{fig.body}, where $h$ and $R_h$ are control parameters for the body height and girth of the human model, while the parameters of other body parts are scaled proportionally to these two values to ensure reasonable body proportions. In this work, the dielectric properties of human tissues have been obtained by the Cole–Cole model \cite{cole-cole}, in which the 6-term model with 14 parameters is used
\begin{equation}\label{eq:cole}   \epsilon_r(\omega)=\epsilon_{r,\infty}+\sum_{i=1}^4\frac{\Delta\epsilon_{r,i}}{1+\left({j}\omega\tau_i\right)^{1-\alpha_i}}-j\frac{\sigma}{\omega\epsilon_0}
\end{equation}
while the model parameters were obtained from Table 1 in \cite{cole-cole}, assuming muscle as a homogeneous human tissue \cite{Dogaru2007muscle,2015TIM}. From the model, at the operating frequency of 2.43 GHz, the material has a complex relative permittivity $\epsilon_r = 52.7-{j}12.76$. 

To validate computational accuracy, both near-field and far-field results are compared with those from Feko$^\text{TM}$ simulations \cite{FEKO2025}. A typical set of human body size parameters is used, with $h = 1.7$ m and $R_h = 0.2$ m. A $z$-directed Hertzian dipole acting as a source term is located at $\mathbf{r}_s = (1, 0, 1)$ m, with power $P_s=1$ W.

\begin{figure}[!t]
\centerline{\includegraphics[width=3in]{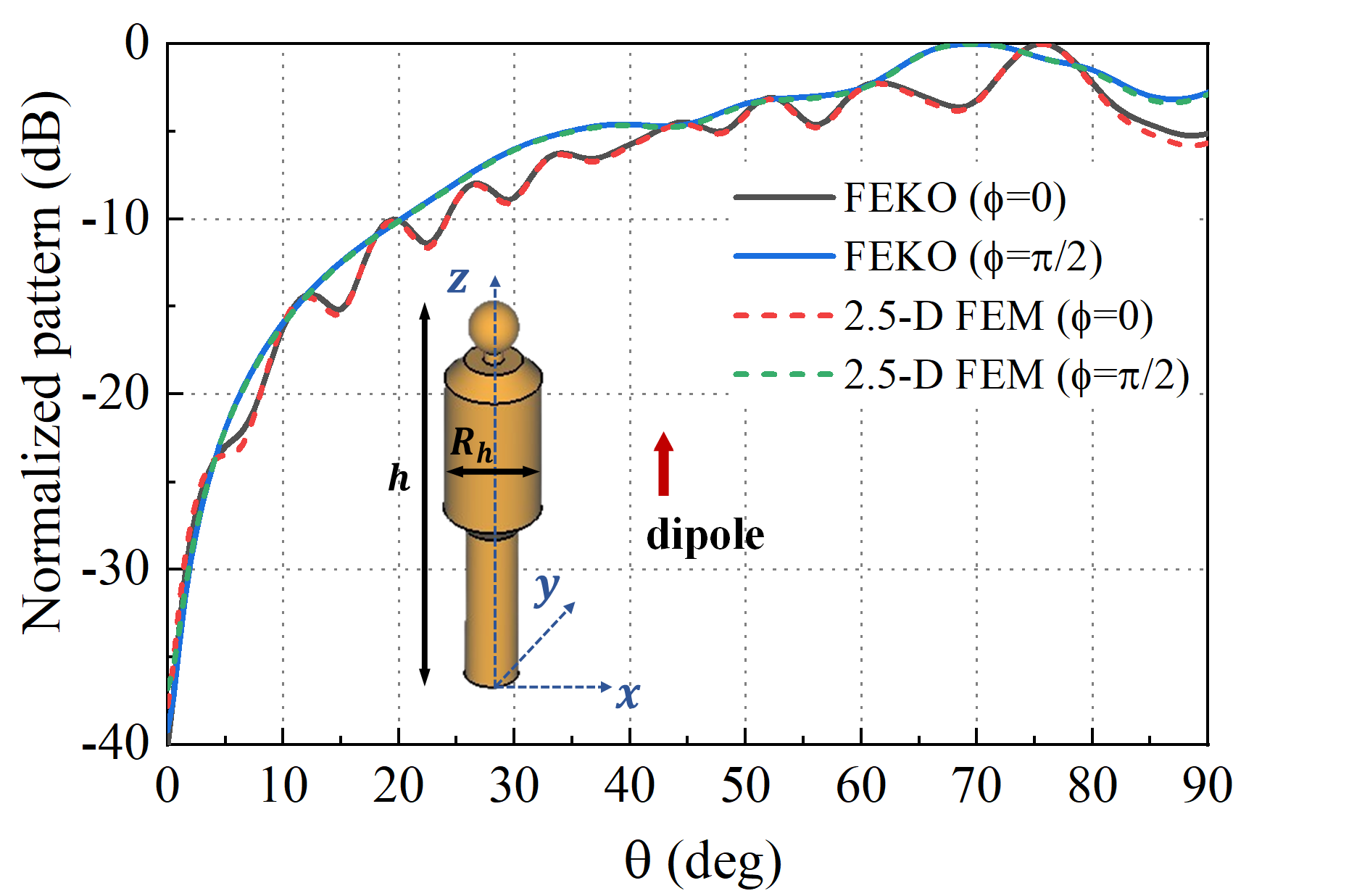}}
\caption{Comparison between 2.5-D FEM results and Feko$^\text{TM}$ simulations: normalized radiation pattern at 2.43 GHz in the $\phi=0$ and $\phi=90^{\circ}$ planes as a function of the coelevation angle $\theta$ for $z$-directed electric dipole in the presence of a BoR human model ($\epsilon_r = 52.7 -j12.76$). The dipole is located at $\mathbf{r}_s = (1, 0, 1)$ m, the BoR human model has height $h = 1.7$ m and radius $R_h = 0.2$ m.}
\label{fig.vali_far}
\end{figure}

\begin{figure}[!t]
\centerline{\includegraphics[width=\columnwidth]{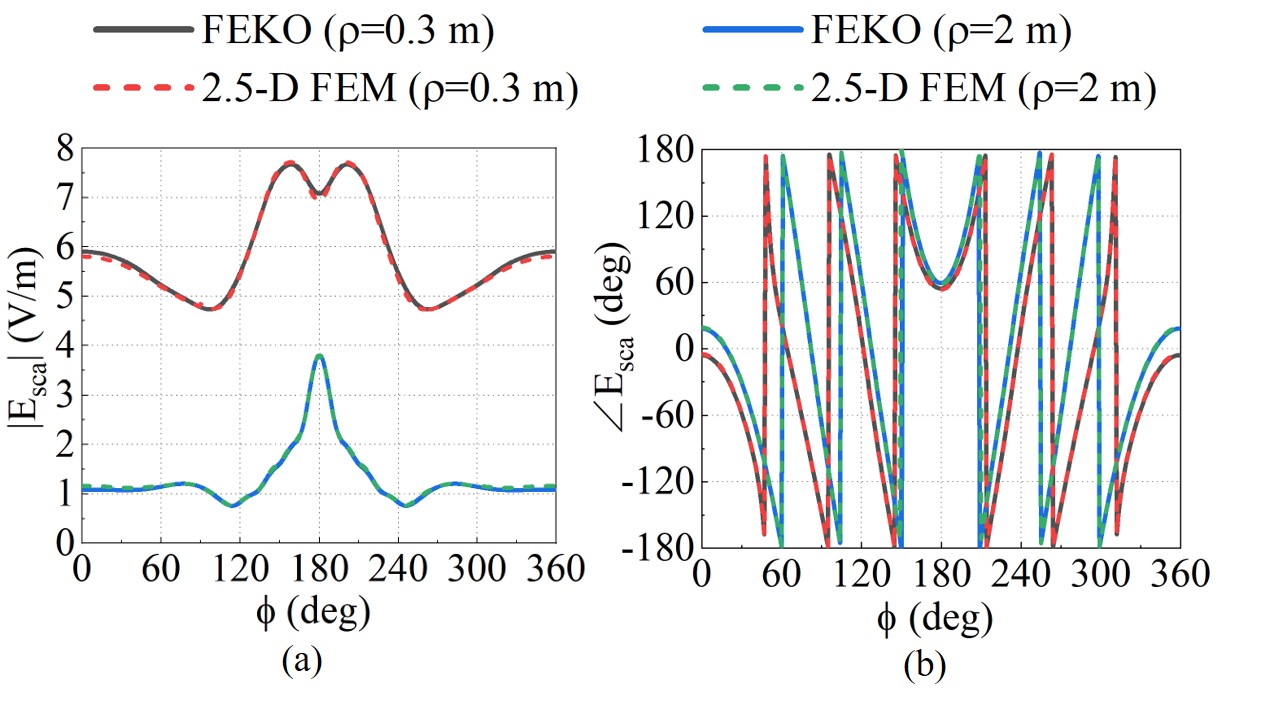}}
\caption{Comparison between 2.5-D FEM results and Feko$^\text{TM}$ simulations of the scattered field ${\bf E}_{\text{sca}}$ at 2.43 GHz along observation points distributed over $\phi\in[0,2\pi]$ for $\rho=0.3$ m and $\rho=2$ m in the presence of a BoR human model ($\epsilon_r = 52.7 -j12.76$): (a) magnitude, and (b) phase. $\phi$ denotes the azimuth angle. The source and BoR parameters are the same as those in Fig.~\ref{fig.vali_far}.}
\label{fig.vali_near}
\end{figure}

Far-field results are presented in Fig.~\ref{fig.vali_far}, where the normalized radiation patterns on the $\phi=0$ and $\phi=90^{\circ}$ planes as a function of $\theta$ are compared with the corresponding Feko$^\text{TM}$ simulations. Here, $\theta$ and $\phi$ denote the coelevation and azimuth angles, respectively, in the local spherical reference frame. Good agreement can be observed between the proposed method and the reference results. Near-field validation is shown in Fig.~\ref{fig.vali_near}. Specifically, the magnitudes and phases of scattered fields are evaluated along observation points distributed over $\phi\in[0,2\pi]$ at $\rho=0.3$ m and $\rho=2$ m, respectively. For the case of $\rho=0.3$ m, the observation distance from the BoR model is smaller than one wavelength, and FEM directly computes the fields. The results show that the average deviation from the results of Feko$^\text{TM}$ is below 0.9\%. For the case of $\rho=2$ m, where the observation points are located several wavelengths away from the BoR model, the fields are reconstructed using the equivalent-source approach. The corresponding average deviation from Feko$^\text{TM}$ results is below 1.2\%, demonstrating the good accuracy of the adopted hybrid strategy. These results validate the accuracy of the proposed 2.5-D FEM from both far-field and near-field perspectives.

\begin{figure*}[!t]
\centerline{\includegraphics[width=\textwidth]{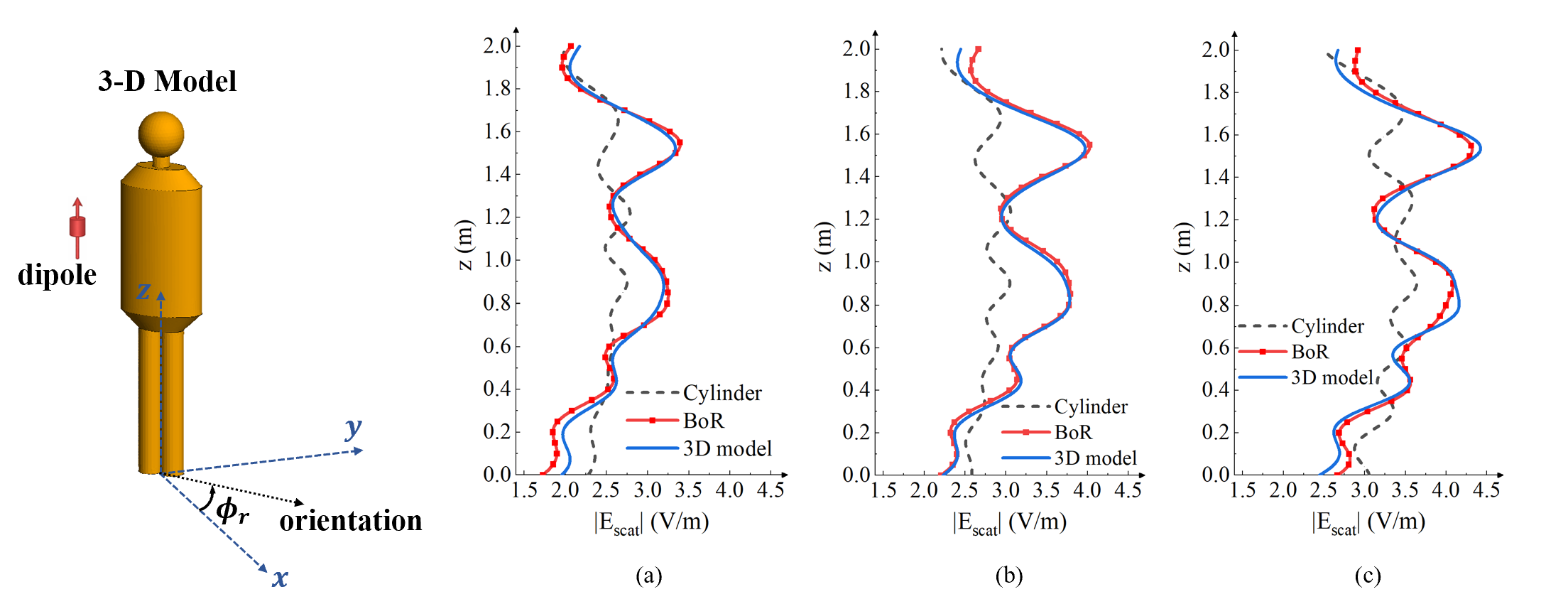}}
\caption{Comparison of scattered field magnitudes for cylindrical, BoR, and 3-D human models. The 3-D model is shown on the left side, while the BoR model is shown in Fig.~\ref{fig.body}. The scattered fields are evaluated along the z-axis at $x=2$ m for a z-directed electric dipole located at $\mathbf{r_s} = (-1, 0, 1)$ m. Results are presented for different body orientations and corresponding equivalent radii of the cylindrical and BoR models (denoted by $R_e$): (a) $\phi_r=0$, $R_e=0.12$ m, (b) $\phi_r=45^{\circ}$, $R_e=0.15$ m, and (c) $\phi_r=90^{\circ}$, $R_e=0.2$ m.}
\label{fig.model_significance}
\end{figure*}

The maximum harmonic index $M$ used in the analysis was 13. The FEM discretization corresponds to 15128 triangles and 106881 degrees of freedom (with 2nd order mixed elements \cite{2.5DFEM04}) for each harmonic in $\phi$. The procedure implemented in MATLAB achieved the result in 37.4 sec with a peak memory usage of 3.1 GB, on a single processor of a PC equipped with an Intel i7 processor working at 2.1 GHz and with 64 GB RAM. In comparison, the total runtime in Feko$^\text{TM}$ was approximately 5744.7 sec, with a peak memory usage of 19.3 GB. This represents a speedup of approximately 154$\times$ and an 84\% reduction in memory consumption compared to the full-wave solution.

\subsection{Assessment of the BoR Human Model}

The proposed BoR human model is to achieve a trade-off between electromagnetic fidelity and computational efficiency for RF sensing applications. Cylindrical approximations provide high computational efficiency but often fail to preserve essential human scattering characteristics due to excessive geometric simplification. Conversely, while realistic 3D models improve accuracy, their computational and memory overhead is prohibitive for large-scale parametric studies and database generation. The BoR model aims to retain the dominant scattering features of the human body while enabling highly efficient full-wave analysis through the 2.5-D FEM. To demonstrate these advantages, we compare these three different models from two perspectives: scattering characteristic preservation and computational cost.

Comparisons are conducted among three different human representations: a simplified cylindrical model, the proposed BoR model, and a 3-D human model. The scattering characteristics of these models are compared by analyzing the magnitudes of scattered fields along the z-axis in a Cartesian coordinate system, over the interval $z\in[0,2]$ m at the observation line $x=2$ m, when a z-directed electric dipole located at $\mathbf{r}_s = (-1, 0, 1)$ m is used as the excitation source.

The adopted 3-D human model is illustrated on the left of Fig.~\ref{fig.model_significance}. The model includes the major anatomical features of the human body, including the head, neck, torso, and legs. In particular, the torso is modeled as an elliptical cylinder to better represent the realistic cross-sectional shape of the human body. According to typical human proportions, the minor axis is chosen as 0.6 times the major axis, with the major and minor radii given by $R_l=0.2$ m and $R_s=0.12$ m, respectively. 

To investigate the influence of the non-axisymmetric characteristics of the human body on the scattering fields, a body orientation angle denoted by $\phi_r$ is introduced as shown in Fig.~\ref{fig.model_significance}. Forward-scattering results are compared for $\phi_r=0$, $\phi_r=45^{\circ}$ and $\phi_r=90^{\circ}$. Here, $\phi_r=0$ corresponds to the case where the dipole is located at the side of the human body, while $\phi_r=90^{\circ}$ represents the case where the dipole is located in front. The BoR and the cylindrical models have the same height as the 3-D model, namely $h=1.7 $ m. Their radii are selected according to the effective cross-sectional radius corresponding to the incident direction of the excitation source.

The scattering results corresponding to different human representations and body orientations are shown in Fig.~\ref{fig.model_significance}. The results for the cylindrical and BoR human models are computed using the proposed 2.5-D FEM, while the results for the 3-D human model are obtained from full-wave simulations using Feko$^\text{TM}$. It can be observed that the proposed BoR model exhibits good agreement with the 3-D human model for all considered body orientations. In particular, the BoR model preserves the dominant variation trend of the scattered fields along the height direction, including the local fluctuations introduced by different anatomical regions. In contrast, the simplified cylindrical model fails to reproduce these spatial variations in scattering. As a result, the cylindrical model produces smoother field distributions and loses important height-dependent scattering characteristics associated with the actual human body structure.

Furthermore, as the effective cross-sectional area of the human body increases with the orientation angle $\phi_r$, the overall magnitude of the scattered fields correspondingly increases, while the variation trend of the scattering fields along the height direction remains consistent. In indoor RF sensing and DFL scenarios, human-body orientation information is generally unavailable due to natural body movements and rotations during measurements. Consequently, the proposed BoR model can be interpreted as an effective orientation-averaged representation of the human body. Accordingly, in the subsequent indoor propagation analysis, the radius of the BoR model is selected as an intermediate value between the major and minor axes of the torso cross-section in order to balance the effects of orientation. The validity of this equivalence is also verified in detail through comparisons with measurements in the following section.

\begin{table}[!t]
\caption{Computational Complexity Comparison of Cylindrical, BoR, and 3-D Human Models}
\label{tab:comparison}
\centering
\renewcommand{\arraystretch}{1.35}
\begin{tabular}{>{\raggedright}p{2.3cm} >{\centering}p{1.4cm} >{\centering}p{1.4cm} >{\centering\arraybackslash}p{1.6cm}}
\hline\hline
 & \textbf{Cylinder} & \textbf{BoR} & \textbf{3-D Model} \\
\hline
\textbf{Unknowns}          & 109481  & 106881  & 40842   \\
\textbf{Peak Memory}       & 3.7\,GB   & 3.1\,GB   & 12.5\,GB  \\
\textbf{Runtime}           & 39.2\,s   & 37.4\,s   & 7447.8\,s  \\
\hline
\end{tabular}
\end{table}

Table~\ref{tab:comparison} summarizes the computational complexity comparison among the cylindrical, BoR, and 3-D human models. All simulations were performed on a single processor of the same PC equipped with the previously mentioned Intel i7 processor detailed in Section~\ref{subsec:validation}. As expected, the 3-D model requires substantial computational resources due to the full 3-D discretization, resulting in significantly increased memory consumption and simulation time. Although the 3-D model provides the highest geometric fidelity, its computational cost becomes prohibitively high for large-scale RF sensing analysis and database generation.

In comparison, the cylindrical and BoR models exhibit similar computational costs since both are rotationally symmetric structures analyzed using the proposed 2.5-D FEM. This indicates that, compared with the cylindrical model, the proposed BoR model is capable of preserving the dominant scattering characteristics compared to a 3-D human model with good accuracy while introducing almost no additional computational cost. Therefore, the proposed BoR human model provides a practical balance between electromagnetic fidelity and computational efficiency in general cases. Its applicability in realistic indoor environments will be further validated in the next section through comparisons with measurements.

\section{Indoor RF Sensing Modeling and Experimental Validation}
\label{chap:experimental_results}

In this section, an indoor DFL scenario is introduced, with a corresponding measurement setup. In order to characterize an arbitrarily complex indoor scenario, we propose a phase-retrieval method based on a small sample of measurements to account for the propagation effects introduced by multipath due to walls, floor, and ceiling reflections, general environmental clutter, as well as the calibration of the transmitting and receiving nodes. The simulation results obtained by combining the proposed BoR scattering model with the indoor environment model are then compared with measurements. Through validation, the proposed model is demonstrated to provide quite accurate predictions for practical indoor RF sensing applications and to serve as a physics-consistent approach for database generation.

\subsection{Indoor Scenario and Measurement Setup}

\begin{figure}[!t]
\centerline{\includegraphics[width=3.2in]{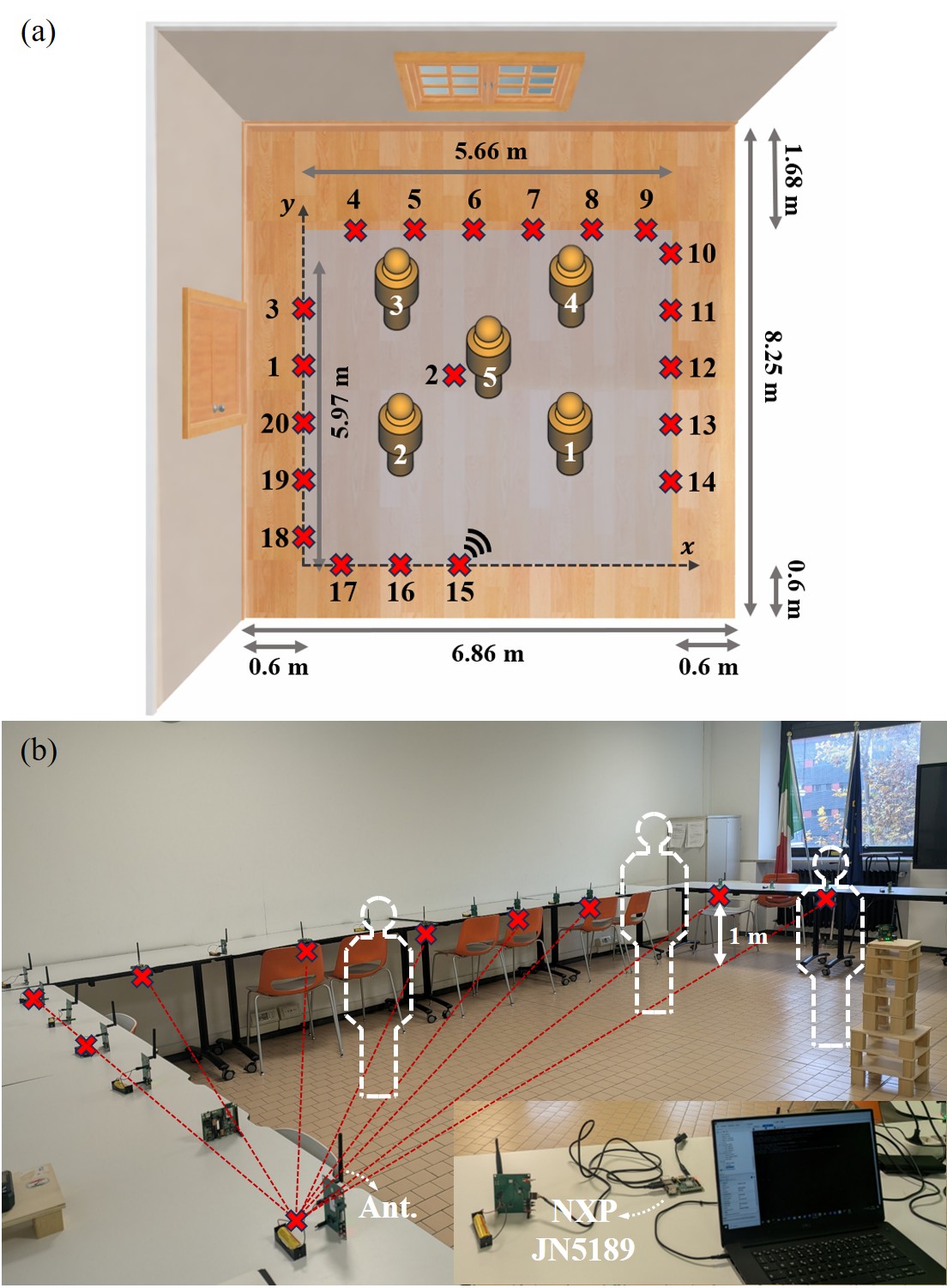}}
\caption{(a) Top-view schematic diagram of the indoor DFL scenario. The gray rectangular region indicates the test area for antennas and human targets. Twenty antennas are deployed as transmitters/receivers, and five human positions are marked. The room entrance is in the lower right corner, where no antennas are deployed. (b) Illustration of the indoor DFL test scenario as seen from the room entrance. The Receiver (Rx)/Transmitter (Tx) deployment is marked by red crosses. Human body positions are reported in dashed white lines. The RF acquisition setup is sketched on the lower right part of this figure.}
\label{fig.sys2}
\end{figure}

The DFL environment is modeled as an indoor space with dimensions illustrated in Fig.~\ref{fig.sys2}(a). Twenty radio transceiver devices (with antennas labeled as $n= 1,\dots, N=20$) are arranged along a rectangular layout within the room, with the antennas uniformly distributed along each side of the rectangle and one antenna (i.e., antenna 2) near the center of the room. Five possible human body locations (labeled as $b = 1,\dots,5$) are considered in the scenario as detailed in the same figure. Taking the lower-left corner of the rectangle as the origin of the $x,y$ plane, the coordinates of the antennas and the predefined human positions are listed in Table~\ref{tab:ant_human_coord}. All antennas are arranged at 1 m above the ground, as a typical indoor RF sensing node arrangement \cite{Federica_SysSetup}.

\begin{table}[t]
\caption{Relative coordinates $(x,y)$ of Antennas and Nominal Locations of the Human Body}
\label{tab:ant_human_coord}
\centering
\renewcommand{\arraystretch}{1.15}
\begin{tabular}{>{\centering\arraybackslash}p{0.65cm} c >{\centering\arraybackslash}p{0.65cm} c >{\centering\arraybackslash}p{0.65cm} c}
\hline\hline
Ant. & Coord. (m) & Ant. & Coord. (m) & Body & Coord. (m) \\
\hline
1  & (0.00, 3.51) & 11 & (5.66, 4.47) & 1 & (4.12, 1.97) \\
2  & (2.85, 3.43) & 12 & (5.66, 3.47) & 2 & (1.48, 1.97) \\
3  & (0.00, 4.51) & 13 & (5.66, 2.47) & 3 & (1.48, 4.61) \\
4  & (0.56, 5.97) & 14 & (5.66, 1.47) & 4 & (4.13, 4.61) \\
5  & (1.56, 5.97) & 15 & (2.50, 0.00) & 5 & (2.65, 3.43) \\
6  & (2.56, 5.97) & 16 & (1.50, 0.00) &   &              \\
7  & (3.56, 5.97) & 17 & (0.50, 0.00) &   &              \\
8  & (4.56, 5.97) & 18 & (0.00, 0.51) &   &              \\
9  & (5.56, 5.97) & 19 & (0.00, 1.51) &   &              \\
10 & (5.66, 5.47) & 20 & (0.00, 2.51) &   &              \\
\hline\hline
\end{tabular}
\end{table}

The corresponding measurement scenario is shown in Fig.~\ref{fig.sys2}(b). The distribution of the RF sensing network nodes is consistent with the simulation scenario described above ~\cite{Federica_SysSetup}. The network uses IoT radio devices operating at 2.43 GHz (channel 16), and it follows the IEEE 802.15.4 physical layer standard, which is commonly adopted in industrial applications~\cite{IEEE802}. The network employs a time-slotted access in which each RF node transmits a physical protocol data unit (beacon) during its assigned time slot, with a slot duration of 0.5 ms and a guard time of 0.15 ms. During each slot, nodes measure the RSSI of all incoming beacons. The collected samples are then sent to a central access point (AP) that serves as a data aggregation and processing hub. Snapshots, consisting of RSSI measurements for all links of the network, are collected every 60 ms. The network is implemented using low-power NXP JN5189 SoC transceivers \cite{NXP_JN5189} and remains fully compatible with the IEEE 802.15.4 standard and common communication protocols such as  RFC 4944 (6LoWPAN) and TSCH (6TiSCH)~\cite{MultiModal}. Each node is equipped with a vertical omni antenna for dual-band 2.4 GHz and 5 GHz, usually employed for WiFi applications.

During the measurements, each $n$-th node is sequentially selected as transmitter, while all remaining $N-1$ nodes (each of them with index $m\ne n=1,\cdots, N$) operate as receivers. For each transmitting configuration, network data are collected using a real human subject positioned at five predefined locations. RSSI measurements $P_{mn,\text{dB}}^{(b)}$, expressed in dBm, are collected by the network nodes as
\begin{equation}\label{eq:P_uv}
    P_{mn,\text{dB}}^{(b)} =
\begin{cases} 
P_{mn,\text{dB}}, & b = 0\text{ (background)} \\
P_{mn,\text{dB}} + \Delta P_{mn,\text{dB}}^{(b)}, & b =1, \cdots,5
\end{cases}
\end{equation}
where $m$ and $n$ denote the transmitting and receiving nodes, respectively, $P_{mn,\text{dB}}$ is the background RSSI (dBm), i.e., with no people inside the monitored area $(b=0)$, which is assumed to be measured in the reference scenario ~\cite{Federica_SysSetup}, and $\Delta P_{mn,\text{dB}}^{(b)}$ is the power variation due to the presence of the human body in the location $b$, a quantity that can be positive or negative.

\subsection{Background-Field Reconstruction Model via Phase Retrieval}

RF sensing prediction requires not only a reasonably accurate computation of human-body scattering but also a proper characterization of the indoor propagation environment. In practical indoor environments, fields are influenced by multipath propagation and reflections from walls, floors, ceilings, and other surrounding scatterers. Moreover, the geometry and EM properties of the environment are often only partially known. As a result, the complex indoor background field cannot be accurately characterized using simple analytical ray-reflection models. Instead of explicitly reconstructing all environmental reflection mechanisms, this work proposes a measurement-driven background-field modeling approach.

Specifically, a phase-retrieval method based on a small sample of measurements, labeled  'training set', is introduced. Starting from the measured background field, the unknown phase information of the background field is recovered by exploiting measurements with human bodies at known positions in the training set. The human body is modeled by a BoR full-wave approach. The result obtained is then verified on another set of measurements with different human positions, labeled 'testing set'.
Letting ${\bf E}_\text{bg}$ be the background field,   ${\bf E}_\text{sca}$ be the field scattered by the human body, and ${\bf E}_\text{tot}$ be the total field, all in the same indoor environment, we can write
\begin{equation}
    {\bf E}_\text{tot} = {\bf E}_\text{bg} + {\bf E}_\text{sca}.
    \label{eq:Esca}
\end{equation}
The receivers sample the $z$-component of the total field and of the background field, but we cannot build an analogue of \eqref{eq:Esca}, since there is no information on the phase of either. The objective of the phase-retrieval procedure is to estimate the unknown background-field phase such that the predicted RSSI amplitudes match the measurements collected with human targets at different positions. Considering the communication link between transmitting node $m$ and receiving node $n$ when a human body is present at position $b$, and letting $\phi_{mn}$ be the corresponding unknown phase of  the background field, the phase estimation problem can be formulated as 
\begin{equation}\label{eq:inversion_problem}
    \phi_{mn}=\arg\min_{\phi}\sum_b[\hat{A}_{mn}^{(b)}(\phi)-A_{mn}^{(b)}]^2,
\end{equation}
where $A_{mn}^{(b)}=[P_{mn}^{(b)}]^{1/2}$ is the field amplitude obtained from the measured RSSI (described in \eqref{eq:P_uv}) when a human target is present at position $b$, and $\hat{A}_{mn}^{(b)}$ denotes the corresponding predicted amplitude, which is approximated as the superposition of the background field and the field scattered by the human target, yielding
\begin{equation}\label{eq:inversion_totalfield}
    \hat{A}_{mn}^{(b)}(\phi) = \left| A_{mn}\, e^{j\phi} + C\, \hat{S}_{mn}^{(b)}\, e^{j\hat{\psi}_{mn}^{(b)}} \right|,
\end{equation}
where $A_{mn}=\sqrt{P_{mn}}$ is the $z$-component amplitude of the background field obtained from the measured background RSSI, $\phi_{mn}$ is the unknown phase of the background field, which is to be optimized, $\hat{S}_{mn}^{(b)}$ and $\hat{\psi}_{mn}^{(b)}$ represent the magnitude and phase of the $z$-component of the field scattered by the human target in position $b$, computed numerically using the 2.5-D FEM, and $C$ is an unknown global calibration constant, jointly optimized with the background-field phases, that converts the unit of electric field into the square root of power while accounting for uncertainties in the transmitter and receiver calibration. Here and throughout the section, quantities marked with a hat $(\hat{\cdot})$ denote predicted values by computation, whereas the corresponding quantities without a hat denote measured values. The BoR model uses average parameters to model possible unknown rotations of the human body in the experiment.

Two assumptions are adopted here in the calculation of the scattered field in \eqref{eq:inversion_totalfield}:
\begin{itemize}
    \item the mutual coupling between the human body and the surrounding indoor scatterers (i.e., the higher-order interactions in which the body-scattered field is reflected by the environment and re-scattered by the body) is neglected;
    \item the excitation is assumed to be the free-space field radiated by the transmitting antenna rather than the actual multipath background field in the calculation of scattered fields of the human body using the proposed 2.5-D FEM.
\end{itemize}
These higher-order contributions are assumed to have a minor impact on the total field.

By expanding the squared magnitude of \eqref{eq:inversion_totalfield}, and with a $C$ fixed at its current estimate, each target position $b$ provides one scalar constraint on the unknown phase $\phi$ as
\begin{equation}\label{eq:inversion_constraint}
    \cos\!\left(\phi-\hat{\psi}_{mn}^{(b)}\right) =
    \frac{(A_{mn}^{(b)})^2 - A_{mn}^2 - (C\hat{S}_{mn}^{(b)})^2}
         {2 A_{mn} C \hat{S}_{mn}^{(b)}}.
\end{equation}
Defining the right-hand side of \eqref{eq:inversion_constraint} as $d_{mn}^{(b)}$, the phase $\phi_{mn}$ is retrieved by minimizing the residual 
\begin{equation}\label{eq:inversion_cost}
  \phi_{mn} = \underset{\phi \in [0, \pi)}{\arg\min} \sum_{b} \left[ \cos(\phi - \hat{\psi}_{mn}^{(b)}) - d_{mn}^{(b)} \right]^2,
\end{equation}
where $b$ is the collection of human positions belonging to the training set.

\subsection{Validation of the Phase-Retrieval Model}

As anticipated, in order to evaluate the capability of the proposed model in unknown scenarios, the five human positions are divided into two non-overlapping sets: a training set and a testing set. This strict separation ensures that the test results provide a realistic assessment of model predictability in new scenarios. To achieve phase reconstruction in \eqref{eq:inversion_cost} with minimal measurement cost, only the measurements corresponding to a single human position are used for training.

The network consists of $N(N-1)=380$ valid unidirectional communication links. The unknown parameters to be estimated include the background-field phases $\{\phi_{mn}\}$ associated with each of the 380 links, together with a global calibration scalar $C$, resulting in a total of 381 unknowns. Since $\{\phi_{mn}\}$ and $C$ are mutually coupled, an alternating iterative optimization strategy is adopted, in which the joint optimization problem is decomposed into two subproblems that are solved alternately. Specifically, all $\phi_{mn}$ are updated while keeping $C$ fixed, followed by updating $C$ while holding all $\phi_{mn}$ fixed. This procedure is repeated until convergence is achieved.

The prediction accuracy of the model is evaluated using the normalized root-mean-square error (NRMSE). The error is computed in the linear  domain, normalized by the corresponding measured amplitudes, and then expressed in dB as
\begin{equation}\label{eq:inversion_RMSE}
\mathrm{NRMSE}_{\mathrm{dB}}=10\log_{10}\left(\frac{\sum_{m,n,b}(\hat{A}_{mn}^{(b)}-A_{mn}^{(b)})^2}{\sum_{m,n,b}
\left(A_{mn}^{(b)}\right)^2}\right),
\end{equation}
where the $b$ is limited to the testing set. The error is computed in the linear domain to avoid the distortion introduced by logarithmic scaling when the signal amplitude becomes very small. In such cases, a small absolute prediction error may be amplified into a large error in dB, causing the NRMSE to be dominated by a few weak-signal links and therefore failing to objectively reflect the overall prediction quality. A negative NRMSE value in dB indicates that the root-mean-square (RMS) prediction error is smaller than the RMS signal amplitude.

To better illustrate the contribution of phase reconstruction to the prediction accuracy, a no-phase model is introduced as a baseline model that does not take into account any phase info. In this model, the body-induced variations $\Delta\hat{P}_{mn}^{(b)}$ is computed from the 2.5-D FEM, and superimposed to the measured background field as 
\begin{equation}\label{eq:no_phase}
    [(\tilde{A}_{mn}^{(b)})^2]_{\text{dB}}=\left[(A_{mn})^2\right]_{\text{dB}}+\Delta\hat{P}_{mn,\text{dB}}^{(b)},
\end{equation}
where $[(\tilde{A}^{(b)}_{mn})^2]_\text{dB}$ denotes the predicted RSSI in dBm of the no-phase model and 
\begin{equation}
\Delta\hat{P}_{mn,\text{dB}}^{(b)}=10\log_{10}\frac{|E_{mn}^{\text{dir}}+E_{mn}^{(b),\text{sca}}|^2}{|E_{mn}^{\text{dir}}|^2},    
\end{equation}
$E_{mn}^{\text{dir}}$ and $E_{mn}^{(b),\text{sca}}$ represent the $z$-component (the only component sampled by the receiving antennas) of the complex direct field and the complex scattered field with human in position $b$, respectively. This baseline represents the straightforward prediction strategy when the phase of the background field is unavailable, as shown in \cite{Vittorio202201}. The predicted values are converted to the linear amplitude domain, and the same normalized error metric as in \eqref{eq:inversion_RMSE} is used to evaluate the baseline.

\begin{figure}[!t]
\centerline{\includegraphics[width=2.8in]{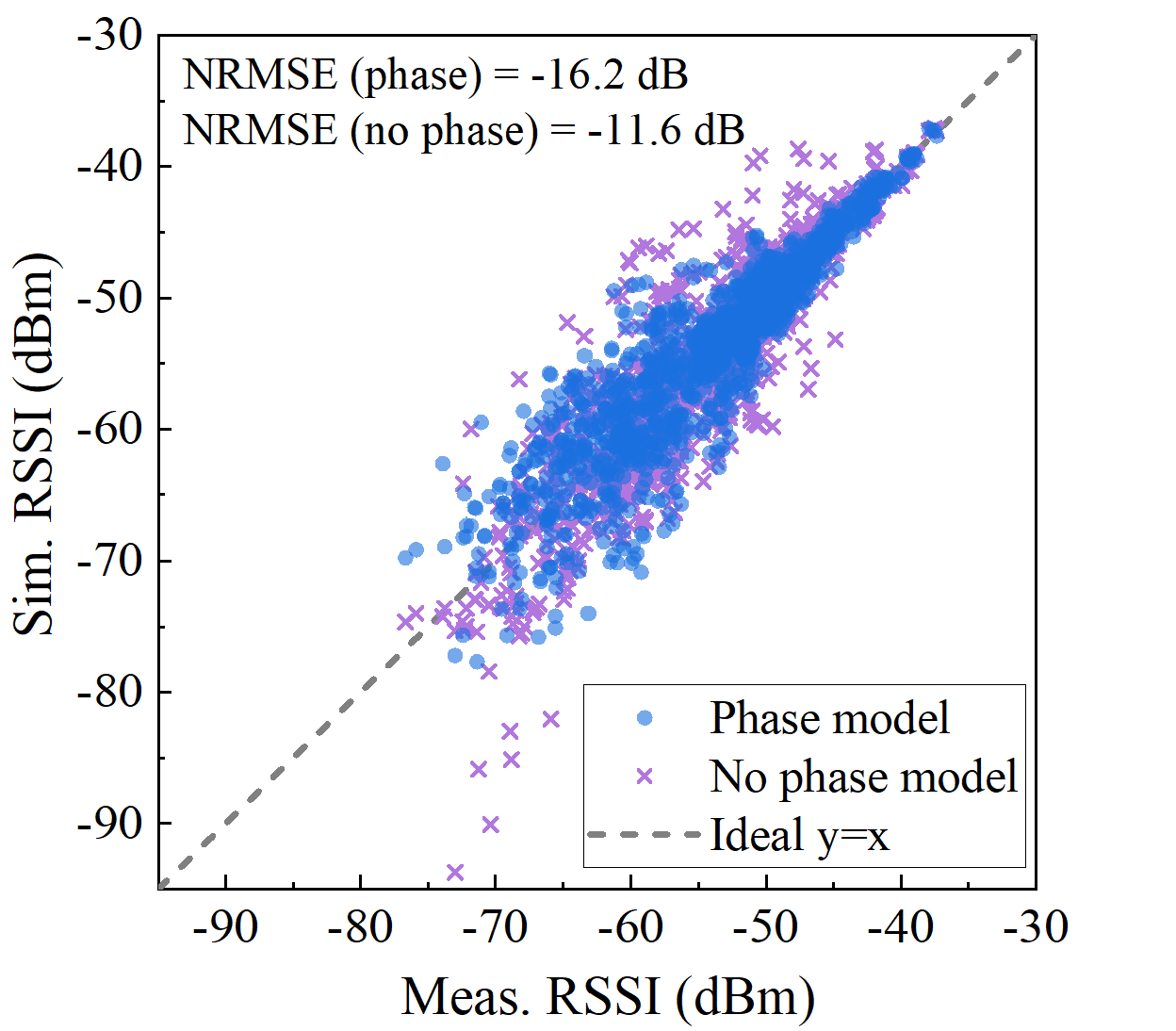}}
\caption{Comparison of predicted and measured RSSI for the phase-retrieval model and the no-phase baseline. Training set position: $b=5$. Test set positions: $b=1\sim4$.}
\label{fig.invers_error}
\end{figure}

Fig.~\ref{fig.invers_error} presents a scatter plot of the predicted RSSI versus the measured RSSI for all valid links at the four test positions (testing-set), using the measurements with the human body located at the predefined position $b=5$ as the training-set. The results of the no-phase model are used as a baseline for comparison. Each point in the figure corresponds to a specific link-position combination. The dashed line represents the ideal case where the predicted and measured RSSI values are identical. Points away from the dashed line indicate a mismatch between measured and predicted RSSI. As shown in the figure, using the estimated phase for the background field produces results that tend to follow the ideal curve better. More specifically, the proposed phase-retrieval model achieves an NRMSE of $-16.2$ dB, corresponding to a normalized prediction error of approximately 15.5\% in linear scale. In contrast, the no-phase model yields an NRMSE of approximately $-11.6$ dB, equivalent to a normalized error of about 26.3\%. These results demonstrate that the optimized background-field phases reduce the relative prediction error by about 35\%, illustrating the effect of phase information. The resulting prediction accuracy is considered acceptable for several RF sensing applications, such as human presence detection and localization.

\begin{figure*}[!t]
\centerline{\includegraphics[width=\textwidth]{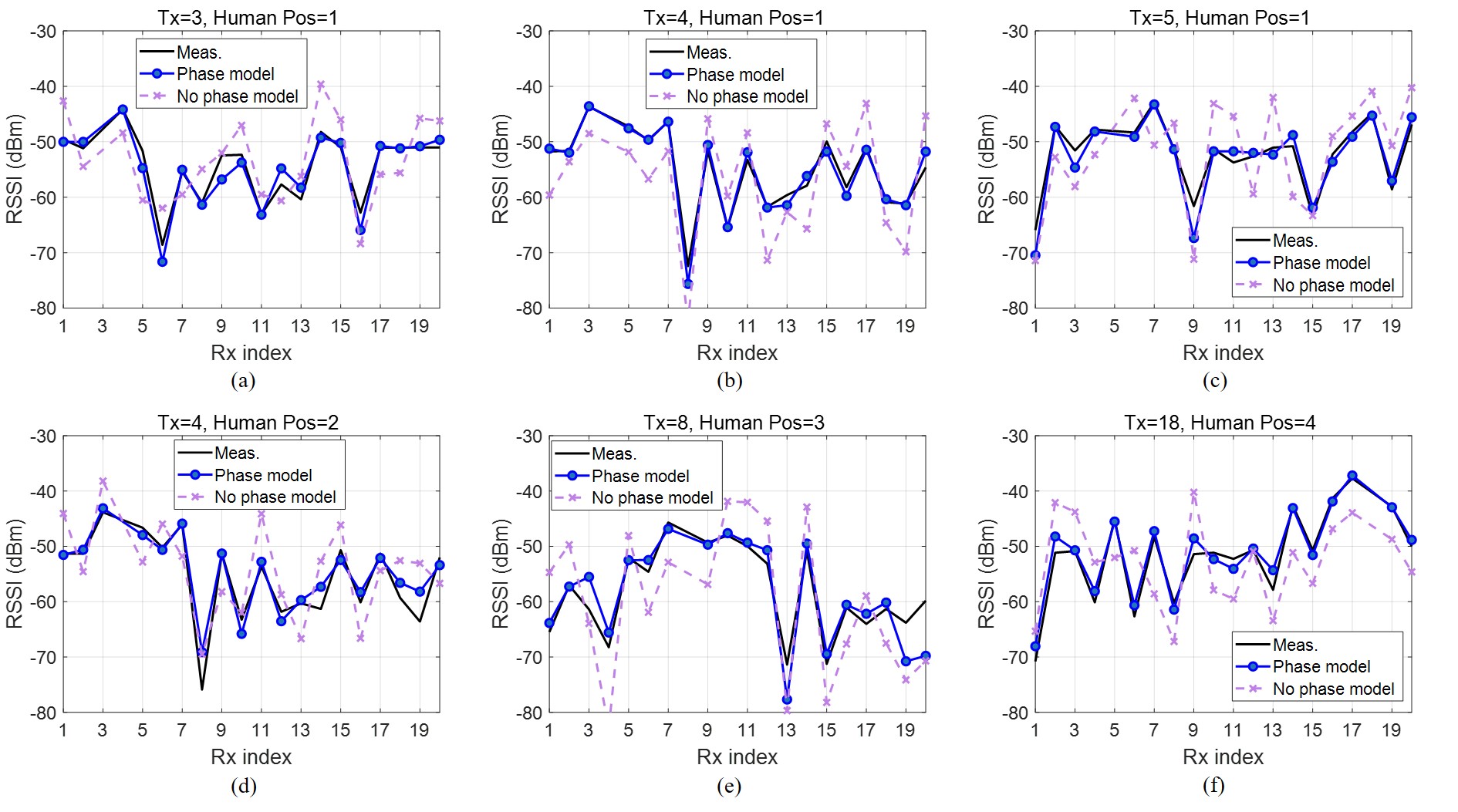}}
\caption{Comparison between measured and predicted RSSI values for several examples of links: (a) Tx = 3, $b=1$, (b) Tx = 4, $b=1$, (c) Tx = 5, $b=1$, (d) Tx = 4, $b=2$, (e) Tx = 8, $b=3$, and (f) Tx = 18, $b=4$. The results of the no-phase model are also included as a baseline for comparison.}
\label{fig.RSSI_link}
\end{figure*}

It is worth noting that the model is trained using measurements from only a single human position $b=5$ and subsequently evaluated on the remaining four positions. The good agreement obtained across all test positions supports the assumptions used in the modelization, i.e., weak contribution of multipath for human body scattered field computation. Moreover, the proposed approach requires only background measurements and calibration data collected at a single human position to reconstruct the background-field information, enabling prediction of the fields at receiving nodes for arbitrary human locations within the environment.

\subsection{Critical comparison with measurements}

The results in Fig.~\ref{fig.invers_error} demonstrate the effectiveness of the proposed measurement-driven phase-retrieval approach from a statistical perspective. In this part, a more detailed validation is conducted at the link level to assess the predictive capability of the proposed model. Since humans were allowed to rotate during the measurements, the girth parameter of the equivalent BoR model in the calculation is chosen as the average of the major and minor body dimensions of the corresponding 3-D human model, that is $R_h=0.16$ m, providing an approximate representation of the random orientation changes occurring in the experiment.

Fig.~\ref{fig.RSSI_link} compares the predicted and measured RSSI for several combinations of transmitting nodes and human positions. In each case, the transmitting node and human position are fixed, while the RSSI values corresponding to all receiving nodes are presented and compared. Since each receiving node experiences a different combination of propagation paths and human-body scattering contributions, these results provide a detailed assessment of the ability of the proposed model to capture the link-dependent characteristics of indoor RF sensing channels. As shown in the figure, the proposed phase-retrieval model exhibits good agreement with the measured results and successfully predicts the RSSI variations across different receiving nodes. More specifically, the proposed model is able to reproduce pronounced local extrema. Examples can be observed in Fig.~\ref{fig.RSSI_link}(d) around Rx = 8 and in Fig.~\ref{fig.RSSI_link}(e) around Rx = 13, where rapid RSSI variations occur due to constructive and destructive interference between the background field and the scattered field. The good agreement between the predicted and measured results in these challenging cases indicates that the reconstructed background-field information successfully captures the dominant propagation characteristics of the indoor environment.

The results of the no-phase model are also presented as a baseline for comparison. Compared with the proposed phase-retrieval model, the no-phase model generally exhibits larger deviations from the measurements, which confirms that simply superimposing the $\Delta\text{RSSI}$ from scattering onto the measured background RSSI is insufficient to accurately describe the interaction between the background field and the scattered field. Therefore, the phase information reconstructed plays an important role in achieving better link-level predictions.

\begin{figure*}[!t]
\centerline{\includegraphics[width=\textwidth]{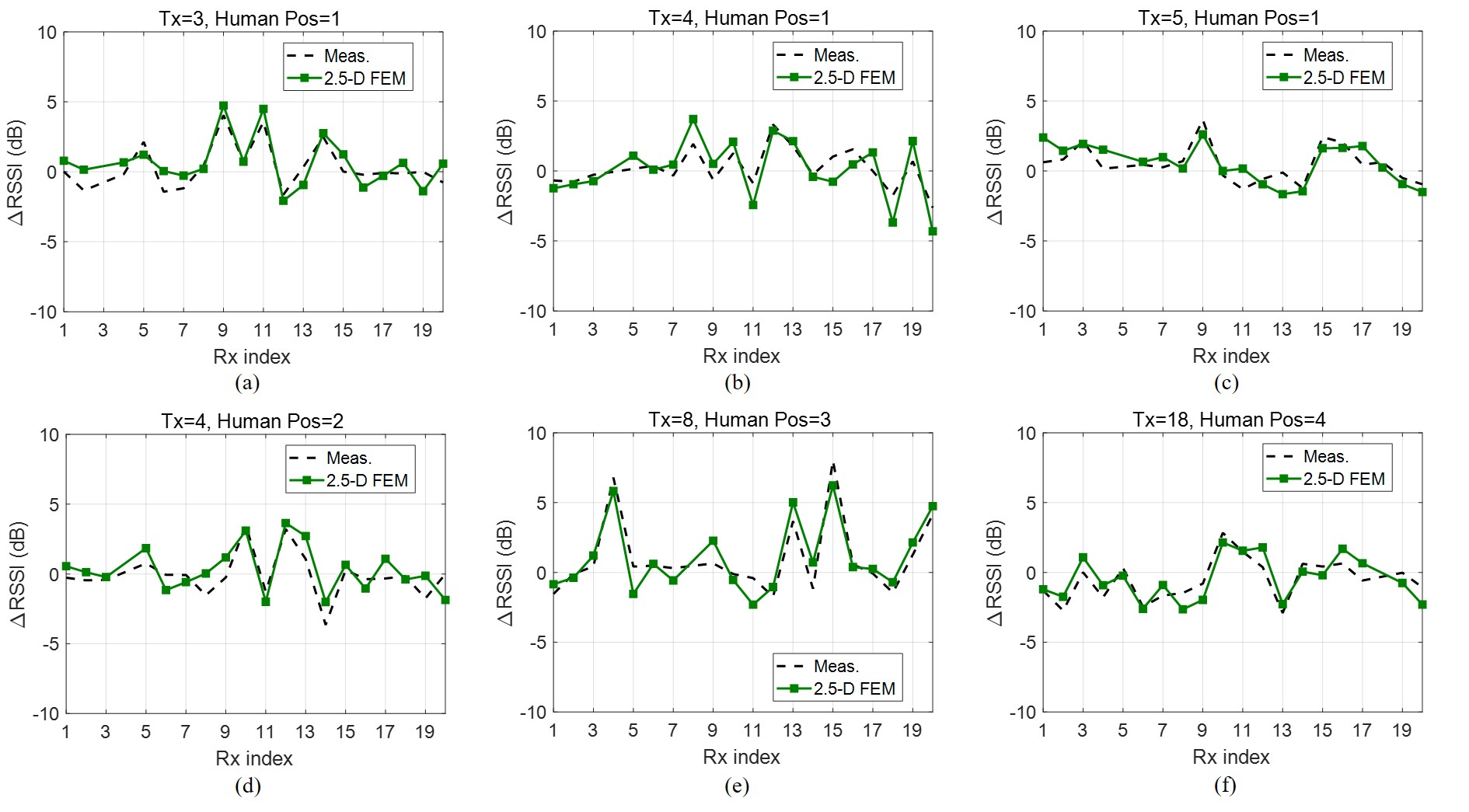}}
\caption{Comparison between measured and predicted $\Delta\text{RSSI}$ values for same examples of links as in Fig.~\ref{fig.RSSI_link}: (a) Tx = 3, $b=1$, (b) Tx = 4, $b=1$, (c) Tx = 5, $b=1$, (d) Tx = 4, $b=2$, (e) Tx = 8, $b=3$, and (f) Tx = 18, $b=4$. The results of 2.5-D FEM are calibrated by an optimized global calibration constant $C$.}
\label{fig.deltaRSSI_link}
\end{figure*}

While the overall prediction capability of the proposed model has been validated, the comparison in Fig.~\ref{fig.deltaRSSI_link} focuses specifically on the human-induced RSSI variations ($\Delta\text{RSSI}$), thereby providing a more direct assessment of the accuracy of the proposed BoR scattering model. For a more direct comparison, the same links presented in Fig.~\ref{fig.RSSI_link} are selected to compare the measured $\Delta\text{RSSI}$ values with those computed by the proposed 2.5-D FEM applied to the BoR human model. The numerical results are calibrated by the optimized global calibration coefficient $C$ obtained through the optimization procedure.

Good agreement can be observed between the simulated and measured results. The proposed 2.5-D FEM and BoR human model successfully reproduces both the magnitude and the spatial variations of the $\Delta\text{RSSI}$ across different receiving nodes. In particular, the locations of local maxima and minima are captured with high accuracy, indicating that the BoR model preserves the dominant scattering mechanisms associated with different body parts and their interactions with the incident EM field. Since the absolute values of $\Delta\text{RSSI}$ are typically within only a few decibels, even relatively small modeling errors may become noticeable in this representation. Nevertheless, the proposed model remains capable of reproducing the overall behavior observed in the measurements. 

In conclusion, the results presented in this section provide a critical validation of the complete model. The proposed measurement-driven environment model successfully incorporates complex indoor propagation effects without requiring explicit reconstruction of multipath mechanisms, while the BoR human model combined with the 2.5-D FEM solver effectively reproduces the scattering behavior observed in measurements. The good agreement obtained for both RSSI and $\Delta\text{RSSI}$ predictions confirms that the proposed BoR model preserves the dominant EM characteristics of the human body while maintaining the computational efficiency required for large-scale simulations, which offers an effective compromise between electromagnetic fidelity and computational cost, making it well-suited for RF sensing prediction and physics-consistent database generation.

\section{Conclusion}
\label{chap:conclusion}

This paper presents a prediction approach for RF sensing applications by combining a BoR human model, an efficient 2.5-D FEM, and a measurement-driven indoor environment modeling approach, which enables rapid prediction of human-body scattering and its impact on wireless channels. The human body is represented as a BoR structure composed of the head, neck, shoulder, torso, hip, and legs. A fast 2.5-D FEM is introduced for computing the scattering fields of the BoR human model under the excitation of a z-directed dipole. The accuracy of the numerical method is comprehensively validated through comparisons with Feko$^\text{TM}$ simulations in both the near-field and far-field regions. Subsequently, comparisons with a simplified cylindrical model and a 3-D human model demonstrate that the proposed BoR representation preserves the dominant scattering features of the 3-D human model while introducing virtually no additional computational cost compared with the cylindrical approximation, thereby achieving a trade-off between electromagnetic fidelity and computational efficiency. To enable practical deployment in indoor DFL scenarios, a measurement-driven background-field reconstruction approach is further introduced. The method utilizes background-field amplitudes from background measurements and reconstructs the corresponding background-field phases from a small set of measurements with targets. As a result, complex propagation effects, including multipath propagation, walls, floor and ceiling reflections, and environmental clutter, are implicitly incorporated into the model without requiring explicit reconstruction of the complex indoor propagation environment. Finally, the proposed approach is systematically validated in a real indoor DFL scenario. By comparing both the RSSI and the human-induced $\Delta\text{RSSI}$ with measurements, good agreement between predictions and measurements is demonstrated from the perspectives of both the total fields and the human-body scattered fields. The results show that the proposed approach not only predicts the influence of the human body on wireless channels but also reproduces the spatial variation patterns observed in measurements. Overall, the proposed approach provides a physically consistent and computationally efficient prediction tool for RF sensing applications. By reducing the reliance on extensive measurement campaigns while generating physically constrained sensing data, the proposed approach offers a practical pathway for the training, evaluation, and database generation of future machine-learning-based DFL and RF sensing systems.

Future work will focus on extending the proposed 2.5-D FEM in two directions. First, the method will be further developed for directional antennas, aiming to enhance the realism of computed scenarios. Second, we will investigate the extension of the method to multi-target environments, establishing interaction models among multiple human bodies to strengthen the capability for more complex and realistic RF simulations of the tool. These developments are expected to enable more comprehensive studies of EM interactions in complex indoor environments.

\appendices
\section{FEM Discretization and Matrix Elements}
\label{app:fem}

From weak-form in \eqref{eq:weak_form}, we can write a general formula in the local cylindrical reference frame as
\begin{equation}
    \int_V (.) \, dV = \int_0^{2\pi}\!\!d\phi\,\!\!\int_S(.)\,\rho \, d\rho \, dz   ,
\end{equation}
where $S$ is a cut plane described by the coordinates $\rho,z$. The integral in $\phi$ can be obtained thanks to the rotational symmetry of 2.5-D FEM, so that the FEM integrals are then limited to $S$ only.

Galerkin's method and a mixed finite-element basis are used to discretize \eqref{eq:weak_form}. Using edge-based basis functions ${\bm\tau}_q(\rho,z)$ to expand the in-plane component ${\bf e}_t^{(m)}$, and Lagrange nodal basis functions $\varphi_q(\rho,z)$ to expand the out-of-plane component ${\rho e_\phi^{(m)}}$, the two components of \eqref{eq:Et_Ephi} can be expressed as
\begin{equation}\label{eq:Et}
    {\bf e}_t^{(m)}(\rho,z) = \sum_{q=1}^{Q} u_{q}^{(m)}{\bm\tau}_q(\rho,z), 
\end{equation}
\begin{equation}\label{eq:Ephi}
    \rho e_\phi^{(m)}(\rho,z) =  \sum_{q=1}^{Q'}v_{q}^{(m)}\varphi_{q}(\rho,z),
\end{equation}
where $Q$ and $Q'$ denote the numbers of basis functions associated with ${\bf e}_t^{(m)}$ and $\rho e_\phi^{(m)}$, respectively. 

Using the subscript '$t$' for the in-plane component and '$\phi$' for the out-of-plane component, from the expansions (\ref{eq:Et})-(\ref{eq:Ephi}), we can verify that
\begin{align}
    & (\nabla\times{\bm\tau}c_m)_t = -\frac{m}{\rho}\left(\tau_z\irho - \tau_\rho \iz\right) s_m,
    \label{eq:curl_vec_t}\\
    & (\nabla\times{\bm\tau}c_m)_\phi = \left(\frac{\partial\tau_\rho}{\partial z}-\frac{\partial \tau_z}{\partial \rho}\right)c_m ,
    \label{eq:curl_vec_p}    
\end{align}
\begin{equation}\label{eq:curl_sca}
    \nabla \times \left(\iphi\frac{\varphi}{\rho }s_m\right)= 
\frac{(\nabla \times \iphi\varphi)_t}{\rho} s_m,
\end{equation}
\begin{equation}
    (\nabla\times\iphi\varphi)_t = 
    \frac{\partial\varphi}{\partial\rho}\iz - \frac{\partial\varphi}{\partial z}\irho.
\end{equation}

In Galerkin's approach, the set of testing functions is the same set of basis functions, so, in addition to the full set of expansion functions
\begin{align}
  {\bf E}_t\rightarrow  \left\{{\bm\tau}_q c_m\right\},\hspace{1em}m=0\ldots M,\hspace{1em} q=1\ldots Q \\
\rho{E}_\phi\rightarrow    \left\{{\varphi}_q s_m\right\},\hspace{1em}m=1\ldots M,\hspace{1em}q=1\ldots Q',
\end{align}
we introduce the sets of testing functions
\begin{align}
  {\bf W}_t\rightarrow  \left\{{\bm\tau}_p c_n\right\},\hspace{1em}n=0\ldots M,\hspace{1em} p=1\ldots Q \\
\rho{W}_\phi\rightarrow    \left\{{\varphi}_q s_n\right\},\hspace{1em}n=1\ldots M,\hspace{1em}p=1\ldots Q',
\end{align}
and we test the weak form using the aforementioned set. We obtain the following sparse system of equations
\begin{equation}\label{eq:main}
    \left(\mathbb{A}-k_0^2\mathbb{B} \right)\mathbb{U} = k_0^2\mathbb{K},
\end{equation}
where $\mathbb{A}$ is block-diagonal 
\begin{equation}
\mathbb{A} =
\begin{bmatrix}
\mathbb{A}^{(0)} & 0 & \cdots & 0 \\
0 & \mathbb{A}^{(1)} & \cdots & 0 \\
\vdots & \vdots & \ddots & \vdots \\
0 & 0 & \cdots & \mathbb{A}^{(M)}
\end{bmatrix},
\end{equation}
$\mathbb{B}=\mathbb{I}\otimes\mathbb{B}'$  with $\otimes$ being the Kronecker product, $\mathbb{I}$ the identity matrix of size $M+1$ and
\begin{equation}
    \mathbb{U} =     
    \begin{bmatrix}
    \mathbb{U}^{(0)} \\
    \mathbb{U}^{(1)} \\
    \vdots   \\
    \mathbb{U}^{(M)}\\
    \end{bmatrix}, \hspace{2em}
    \mathbb{K} =     
    \begin{bmatrix}
    \mathbb{K}^{(0)} \\
    \mathbb{K}^{(1)} \\
    \vdots   \\
    \mathbb{K}^{(M)}\\
    \end{bmatrix}.
\end{equation}

The block-diagonal structure of the system in \eqref{eq:main} is due to the fact that
\begin{equation}
    \int_0^{2\pi} c_m c_n d\phi = 
    \left\{
    \begin{array}{ll}
    1 &\text{if }m=n \\
    0 &\text{if }m\neq n \\    
    \end{array}
    \right. ,
\end{equation}
\begin{equation}
    \int_0^{2\pi} s_m s_n d\phi = 
    \left\{
    \begin{array}{ll}
    1 &\text{if }m=n>0 \\
    0 &\text{if }m\neq n \\    
    \end{array}
    \right. ,
\end{equation}
\begin{equation}
    \int_0^{2\pi} c_m s_n d\phi = 0.
\end{equation}
We can further express the individual matrices of the system as
\begin{equation}\label{eq:Amat}
    \mathbb{A}^{(m)} = \begin{bmatrix}
        \mathbb{A}_{tt} + m^2\mathbb{A}'_{tt} & -m\mathbb{A}_{t\phi} \\
        -m\mathbb{A}_{\phi t} & \mathbb{A}_{\phi\phi} \\       
    \end{bmatrix} ,
\end{equation} 
\begin{equation}\label{eq:Bmat}
\mathbb{B}' =\begin{bmatrix}
         \mathbb{B}_{tt} & {0} \\
        {0} & \mathbb{B}_{\phi\phi} \\       
\end{bmatrix} ,
\end{equation}
\begin{equation}
        \mathbb{U}^{(m)} = \begin{bmatrix}\mathbb{U}_{t}^{(m)} \\ \mathbb{U}_{\phi}^{(m)} 
    \end{bmatrix},\hspace{1em}
\mathbb{K}^{(m)} = \begin{bmatrix}
        \mathbb{K}_{t}^{(m)} \\ \mathbb{K}_{\phi}^{(m)} 
    \end{bmatrix},
    \label{eq:knownterm}
\end{equation}
and 
\begin{equation}
    \mathbb{U}_t^{(m)} = [u_1^{(m)}\; u_2^{(m)} \ldots u_{Q}^{(m)}]^\text{T},
\end{equation} 
\begin{equation}
\mathbb{U}_\phi^{(m)} = [v_1^{(m)}\; v_2^{(m)} \ldots v_{Q'}^{(m)}]^\text{T}.
\end{equation}  
To express the specific terms in \eqref{eq:Amat}-\eqref{eq:knownterm}, we let
\begin{equation}
    (\nabla\times{\bm\tau})_t = {\tau_z\irho -\tau_\rho\iz},
\end{equation}
\begin{equation}
    (\nabla\times{\bm\tau})_\phi = \frac{\partial\tau_{\rho}}{\partial z}-\frac{\partial \tau_{z}}{\partial \rho},
\end{equation}
and we have, indicating with $\mathbb{M}(p,q)$ element $p,q$ of matrix $\mathbb{M}$
\begin{equation}
    \mathbb{A}_{tt}(p,q) = \int_S (\nabla\times{\bm\tau}_p)_\phi\cdot(\nabla\times{\bm\tau}_q)_\phi \;\rho d\rho dz,
\end{equation}
\begin{equation}
    \mathbb{A}'_{tt}(p,q) = \int_S (\nabla\times{\bm\tau}_p)_t\cdot(\nabla\times{\bm\tau}_q)_t \frac{1}{\rho} d\rho dz,
\end{equation}
\begin{equation}
    \mathbb{A}_{t\phi}(p,q) = \int_S (\nabla\times{\bm\tau}_p)_t\cdot(\nabla\times\iphi\varphi_q)_t \frac{1}{\rho} d\rho dz,
\end{equation}
\begin{equation}
    \mathbb{A}_{\phi t}(p,q) = \int_S (\nabla\times\iphi\varphi_p)_t\cdot(\nabla\times{\bm\tau}_q)_t \frac{1}{\rho} d\rho dz,
\end{equation}
\begin{equation}
    \mathbb{A}_{\phi\phi}(p,q) = \int_S (\nabla\times\iphi\varphi_p)_t\cdot(\nabla\times\iphi\varphi_q)_t \frac{1}{\rho} d\rho dz,
\end{equation}
\begin{equation}
    \mathbb{B}_{tt}(p,q) = \int_S \epsilon_r {\bm\tau}_p\cdot {\bm\tau}_q \;\rho d\rho dz,
\end{equation}
\begin{equation}
    \mathbb{B}_{\phi\phi}(p,q) = \int_S \epsilon_r \frac{{\varphi}_p {\varphi}_q}{\rho} \; d\rho dz.
\end{equation}
Similar expressions hold for the PML regions, in which a diagonal tensor with components $\Lambda_{\rho\rho}$, $\Lambda_{zz}$ and $\Lambda_{\phi\phi}$, and its inverse represent the constitutive parameters representing complex stretched coordinates \cite{Jinbook}. 

\bibliographystyle{ieeetr}
\bibliography{ref}

@article{cole-cole,
doi = {10.1088/0031-9155/41/11/003},
url = {https://doi.org/10.1088/0031-9155/41/11/003},
year = {1996},
month = {nov},
publisher = {},
volume = {41},
number = {11},
pages = {2271-2293},
author = {S Gabriel and R W Lau and C Gabriel},
title = {The dielectric properties of biological tissues: {III}. Parametric models for the dielectric spectrum of tissues},
journal = {Phys. Med. Biol.}
}

@INPROCEEDINGS{Plouhinec2023Cylinder,
  author={Plouhinec, Eric and Uguen, Bernard},
  booktitle={2023 IEEE-APS Topical Conference on Antennas and Propagation in Wireless Communications (APWC)}, 
  title={A UTD Elliptic Cylinder Model for Studying Body Orientation Influence on Human Blockage}, 
  year={2023},
  volume={},
  number={},
  pages={068-073},
  doi={10.1109/APWC57320.2023.10297474}}

@book{berenger2007perfectly,
  title={Perfectly matched layer (PML) for computational electromagnetics},
  author={B{\'e}renger, Jean-Pierre},
  volume={8},
  year={2007},
  publisher={Springer}
}

@ARTICLE{DFL_healthcare,
  author={Petrov, Vitaly and Mikhaylov, Konstantin and Moltchanov, Dmitri and Andreev, Sergey and Fodor, Gabor and Torsner, Johan and Yanikomeroglu, Halim and Juntti, Markku and Koucheryavy, Yevgeni},
  journal={IEEE Commun. Mag.}, 
  title={When {IoT} Keeps People in the Loop: A Path Towards a New Global Utility}, 
  year={2019},
  volume={57},
  number={1},
  pages={114-121},
  doi={10.1109/MCOM.2018.1700018}}

@article{DFL_smartcity,
title = {A cloud-based architecture for emergency management and first responders localization in smart city environments},
journal = {Comput. Electr. Eng.},
volume = {56},
pages = {810-830},
year = {2016},
issn = {0045-7906},
doi = {https://doi.org/10.1016/j.compeleceng.2016.02.012},
url = {https://www.sciencedirect.com/science/article/pii/S0045790616300283},
author = {Francesco Palmieri and Massimo Ficco and Silvio Pardi and Aniello Castiglione}
}

@ARTICLE{DFL_HumanActive,
  author={Shit, Rathin Chandra and Sharma, Suraj and Puthal, Deepak and James, Philip and Pradhan, Biswajeet and Moorsel, Aad van and Zomaya, Albert Y. and Ranjan, Rajiv},
  journal={IEEE Commun. Surv. Tutor.}, 
  title={Ubiquitous Localization (UbiLoc): A Survey and Taxonomy on Device Free Localization for Smart World}, 
  year={2019},
  volume={21},
  number={4},
  pages={3532-3564},
  doi={10.1109/COMST.2019.2915923}}

@ARTICLE{Vittorio202201,
  author={Rampa, Vittorio and Gentili, Gian Guido and Savazzi, Stefano and D’Amico, Michele},
  journal={IEEE Trans. Antennas Propag.}, 
  title={Electromagnetic Models for Passive Detection and Localization of Multiple Bodies}, 
  year={2022},
  volume={70},
  number={2},
  pages={1462-1475},
  doi={10.1109/TAP.2021.3111405}}

@INPROCEEDINGS{Vittorio202202,
  author={Rampa, Vittorio and Savazzi, Stefano and D’Amico, Michele},
  booktitle={Proc. IEEE-APS Topical Conf. Antennas Propag. Wireless Commun.}, 
  title={Electromagnetic Models for Device-Free Radio Localization with Antenna Arrays}, 
  year={2022},
  volume={},
  number={},
  pages={042-046},
  doi={10.1109/APWC49427.2022.9899896}}

@ARTICLE{Stefano2019,
   author={Savazzi, Stefano and Cammers-Goodwin, Sage and Paulus, Alexander and Kianoush, Sanaz and Golipoor, Sahar and Liu, Ying and Salami, Dariush and Eibert, Thomas and Aydin, Ciano},
  journal={IEEE Communications Magazine}, 
  title={Holography with Dense Wireless Networks: A Case for Ethical Design}, 
  year={2026},
  volume={},
  number={},
  pages={1-7},
  doi={10.1109/MCOM.001.2400733}}

@ARTICLE{Fieramosca_awpl,
  author={Fieramosca, Federica and Rampa, Vittorio and Savazzi, Stefano and D'Amico, Michele},
  journal={IEEE Antennas Wirel. Propag. Lett.}, 
  title={On the Impact of the Antenna Radiation Patterns in Passive Radio Sensing}, 
  year={2024},
  volume={23},
  number={2},
  pages={503-507},
  doi={10.1109/LAWP.2023.3327955}}

@article{review2017,
title = {Recent advances in {RF}-based passive device-free localisation for indoor applications},
journal = {Ad Hoc Netw.},
volume = {64},
pages = {80-98},
year = {2017},
issn = {1570-8705},
doi = {https://doi.org/10.1016/j.adhoc.2017.06.007},
url = {https://www.sciencedirect.com/science/article/pii/S1570870517301257},
author = {Sameera Palipana and Bastien Pietropaoli and Dirk Pesch}
}

@ARTICLE{ML_Bayesian,
  author={Kaltiokallio, Ossi and Hostettler, Roland and Patwari, Neal},
  journal={IEEE Trans. Mob. Comput.}, 
  title={A Novel Bayesian Filter for {RSS}-Based Device-Free Localization and Tracking}, 
  year={2021},
  volume={20},
  number={3},
  pages={780-795},
  doi={10.1109/TMC.2019.2953474}}

@INPROCEEDINGS{ML_VAE_Stefano,
  author={Savazzi, Stefano and Fieramosca, Federica and Kianoush, Sanaz and Rampa, Vittorio and D'Amico, Michele},
  booktitle={Proc. IEEE-APS Topical Conf. Antennas Propag. Wireless Commun.}, 
  title={A Physics-Informed Generative Model for Passive Radio-Frequency Sensing}, 
  year={2023},
  volume={},
  number={},
  pages={056-061},
  doi={10.1109/APWC57320.2023.10297498}}

@ARTICLE{ML_GNN_Stefano,
  author={Savazzi, Stefano and Fieramosca, Federica and Kianoush, Sanaz and D’Amico, Michele and Rampa, Vittorio},
  journal={IEEE Open J. Antennas Propag.}, 
  title={Electromagnetic-Informed Generative Models for Passive RF Sensing and Perception of Body Motions}, 
  year={2024},
  volume={5},
  number={4},
  pages={958-973},
  doi={10.1109/OJAP.2024.3407199}}

@INPROCEEDINGS{ML_VAE_Federica,
  author={Fieramosca, Federica and Rampa, Vittorio and D'Amico, Michele and Savazzi, Stefano},
  booktitle={Proc. 18th Eur. Conf. Antennas Propag. (EuCAP 2024)}, 
  title={Physics-Informed Generative Neural Networks for RF Propagation Prediction with Application to Indoor Body Perception}, 
  year={2024},
  volume={},
  number={},
  pages={1-5},
  doi={10.23919/EuCAP60739.2024.10501077}}

@ARTICLE{ML_GNN_Barba,
  author={Di Barba, Paolo},
  journal={IEEE Trans. Magn.}, 
  title={Future Trends in Optimal Design in Electromagnetics}, 
  year={2022},
  volume={58},
  number={9},
  pages={1-4},
  doi={10.1109/TMAG.2022.3164204}}

@ARTICLE{ML_review,
  author={He, Ruisi and Lau, Buon Kiong and Oestges, Claude and Haneda, Katsuyuki and Liu, Bo},
  journal={IEEE Trans. Antennas Propag.}, 
  title={Guest Editorial Artificial Intelligence in Radio Propagation for Communications}, 
  year={2022},
  volume={70},
  number={6},
  pages={3934-3938},
  doi={10.1109/TAP.2022.3178164}}

@ARTICLE{3DFEM,
  author={Takei, Amane and Murotani, Kohei and Sugimoto, Shin-Ichiro and Ogino, Masao and Kawai, Hiroshi},
  journal={IEEE Trans. Magn.}, 
  title={High-Accuracy Electromagnetic Field Simulation Using Numerical Human Body Models}, 
  year={2016},
  volume={52},
  number={3},
  pages={1-4},
  doi={10.1109/TMAG.2015.2479467}}

@ARTICLE{MOM01,
  author={Keshmiri, Farshad and Craeye, Christophe},
  journal={IEEE Trans. Antennas Propag.}, 
  title={Moment-Method Analysis of Normal-to-Body Antennas Using a {Green's} Function Approach}, 
  year={2012},
  volume={60},
  number={9},
  pages={4259-4270},
  doi={10.1109/TAP.2012.2207053}}

@ARTICLE{MOM02,
  author={Ahmadi, Leila and Shishegar, Amir Ahmad},
  journal={IEEE Trans. Antennas Propag.}, 
  title={A Semi-Explicit Solution of Scattering From Dielectric Bodies Using the Method of Moments}, 
  year={2023},
  volume={71},
  number={8},
  pages={6806-6813},
  doi={10.1109/TAP.2023.3284158}}

@ARTICLE{FDTD01,
  author={Fujii, Katsuyuki and Takahashi, Masaharu and Ito, Koichi},
  journal={IEEE Trans. Antennas Propag.}, 
  title={Electric Field Distributions of Wearable Devices Using the Human Body as a Transmission Channel}, 
  year={2007},
  volume={55},
  number={7},
  pages={2080-2087},
  doi={10.1109/TAP.2007.900226}}

@ARTICLE{FDTD02,
  author={Jurgens, T.G. and Taflove, A.},
  journal={IEEE Trans. Antennas Propag.}, 
  title={Three-dimensional contour {FDTD} modeling of scattering from single and multiple bodies}, 
  year={1993},
  volume={41},
  number={12},
  pages={1703-1708},
  doi={10.1109/8.273315}}

@ARTICLE{GreenHead,
  author={Reyhani, S.M.S. and Ludwig, S.A.},
  journal={IEEE Trans. Biomed. Eng.}, 
  title={An Implanted Spherical Head Model Exposed to Electromagnetic Fields at a Mobile Communication Frequency}, 
  year={2006},
  volume={53},
  number={10},
  pages={2092-2101},
  doi={10.1109/TBME.2006.881770}}

@article{AnaCylind01,
author = {Mahmoud Zamel, Hany and Eldiwany, Essam and El-Hennawy, Hadia},
year = {2014},
month = {01},
pages = {59-70},
title = {Electromagnetic scattering by approximately cloaked dielectric cylinder},
volume = {59},
journal = {Prog. Electromagn. Res. B},
doi = {10.2528/PIERB14011806}
}

@INPROCEEDINGS{AnaCylind02,
  author={Yokota, M. and Ikegamai, T. and Ohta, Y. and Fujii, T.},
  booktitle={Proc. 4th Eur. Conf. Antennas Propag. (EuCAP 2010)}, 
  title={Numerical examination of {EM} wave shadowing by human body}, 
  year={2010},
  volume={},
  number={},
  pages={1-4},
  doi={}}

@article{AnaCylind03,
author = {Awan, Zeeshan A. and Seetharamdoo, Divitha},
title = {Enhancement of backscattering cross section from an equivalent human cylinder under oblique incidence},
journal = {IET Microw. Antennas Propag.},
volume = {16},
number = {8},
pages = {497-509},
doi = {https://doi.org/10.1049/mia2.12256},
url = {https://ietresearch.onlinelibrary.wiley.com/doi/abs/10.1049/mia2.12256},
eprint = {https://ietresearch.onlinelibrary.wiley.com/doi/pdf/10.1049/mia2.12256},
year = {2022}
}

@ARTICLE{hf_method01,
  author={Abdelgawwad, Ahmed and Mallofré, Andreu Català and Pätzold, Matthias},
  journal={IEEE Access}, 
  title={A Trajectory-Driven {3D} Channel Model for Human Activity Recognition}, 
  year={2021},
  volume={9},
  number={},
  pages={103393-103406},
  doi={10.1109/ACCESS.2021.3098951}}

@ARTICLE{hf_method02,
  author={Koutitas, George},
  journal={IEEE Antennas Wirel. Propag. Lett.}, 
  title={Multiple Human Effects in Body Area Networks}, 
  year={2010},
  volume={9},
  number={},
  pages={938-941},
  doi={10.1109/LAWP.2010.2082485}}

@article{2.5DFEM02,
author = {Codecasa, Lorenzo and Gentili, Gian Guido and Khosronejad, Misagh and Pelosi, Giuseppe and Selleri, Stefano},
title = {Exact conic section arc elements in {2D} and {2.5D} {FEM} using a coordinate transformation},
journal = {IET Microw. Antennas Propag.},
volume = {15},
number = {9},
pages = {1108-1116},
doi = {https://doi.org/10.1049/mia2.12107},
url = {https://ietresearch.onlinelibrary.wiley.com/doi/abs/10.1049/mia2.12107},
eprint = {https://ietresearch.onlinelibrary.wiley.com/doi/pdf/10.1049/mia2.12107},
year = {2021}
}

@ARTICLE{2.5DFEM03,
  author={Gentili, Gian Guido and Khosronejad, Misagh and Nesti, Renzo and Pelosi, Giuseppe and Selleri, Stefano},
  journal={IEEE Trans. Antennas Propag.}, 
  title={An Efficient {2.5-D} Finite-Element Approach Based on Transformation Optics for the Analysis of Elliptical Horns}, 
  year={2018},
  volume={66},
  number={9},
  pages={4782-4790},
  doi={10.1109/TAP.2018.2851289}}

@ARTICLE{2.5DFEM04,
  author={Gentili, G. G. and Bolli, P. and Nesti, R. and Pelosi, G. and Toso, L.},
  journal={IEEE Trans. Antennas Propag.}, 
  title={High-Order {FEM} Mode Matching Analysis of Circular Horns With Rotationally Symmetric Dielectrics}, 
  year={2007},
  volume={55},
  number={10},
  pages={2915-2918},
  doi={10.1109/TAP.2007.905956}}

@ARTICLE{IEEE802,
  author={Davoli, Luca and Belli, Laura and Cilfone, Antonio and Ferrari, Gianluigi},
  journal={IEEE Internet Things J.}, 
  title={From Micro to Macro {IoT}: Challenges and Solutions in the Integration of {IEEE} 802.15.4/802.11 and {Sub-GHz} Technologies}, 
  year={2018},
  volume={5},
  number={2},
  pages={784-793},
  doi={10.1109/JIOT.2017.2747900}}

@article{MultiModal,
  title={Multi‐Modal Industrial IoT Networks: Recent Advances and Future Challenges},
  author={Elsas, Robbe and Van Leemput, Dries and Hoebeke, Jeroen and De Poorter, Eli},
  journal={Wirel. Pers. Commun.},
  volume={140},
  pages={1-24},
  year={2025},
  doi={10.1007/s11277-023-10213-w},
  publisher={Springer}
}

@ARTICLE{Federica_SysSetup,
  author={Fieramosca, Federica and Rampa, Vittorio and D'Amico, Michele and Savazzi, Stefano},
  journal={IEEE Internet Things J.}, 
  title={{RF} Sensing With Dense {IoT} Network Graphs: An E{M}-Informed Analysis}, 
  year={2026},
  volume={13},
  number={3},
  pages={5124-5137},
  doi={10.1109/JIOT.2025.3643133}}

@ARTICLE{BOR_FEM,
  author={Zhai, Yong Bo and Ping, Xue Wei and Cui, Tie Jun},
  journal={IEEE Trans. Antennas Propag.}, 
  title={Scattering From Complex Bodies of Revolution Using a High-Order Mixed Finite Element Method and Locally-Conformal Perfectly Matched Layer}, 
  year={2011},
  volume={59},
  number={5},
  pages={1761-1764},
  doi={10.1109/TAP.2011.2122224}}

@ARTICLE{Jinjianming_borfem01,
  author={Greenwood, A.D. and Jian-Ming Jin},
  journal={IEEE Trans. Antennas Propag.}, 
  title={A novel efficient algorithm for scattering from a complex {BOR} using mixed finite elements and cylindrical {PML}}, 
  year={1999},
  volume={47},
  number={4},
  pages={620-629},
  doi={10.1109/8.768800}}

@ARTICLE{Jinjianming_borfem02,
  author={Greenwood, A.D. and Jian-Ming Jin},
  journal={IEEE Trans. Antennas Propag.}, 
  title={Finite-element analysis of complex axisymmetric radiating structures}, 
  year={1999},
  volume={47},
  number={8},
  pages={1260-1266},
  doi={10.1109/8.791941}}

@misc{FEKO2025,
  author = {{Altair Engineering}},
  title = {{FEKO} Electromagnetic Simulation Software},
  howpublished = {Version 2025},
  note = {Altair Engineering GmbH, T{\"u}bingen, Germany},
  url = {https://altair.com/feko/}
}

@misc{NXP_JN5189,
  author       = {{NXP Semiconductors}},
  title        = {{IEEE 802.15.4 Low Power Wireless MCU Rev. 1.3: JN5189 Product Data Sheet}},
  year         = {2021},
  month        = may,
  howpublished = {\url{https://www.nxp.com/docs/en/nxp/data-sheets/JN5189.pdf}}
}

@book{Jinbook,
  author       = {Jian-Ming Jin},
  title        = {The Finite Element Method in Electromagnetics},
  publisher    = {Wiley-IEEE Press},
  location     = {Hoboken, NJ, USA},
  year         = {2014},
  edition      = {3rd},
  isbn         = {9781118571361},
  langid       = {English},
}

@ARTICLE{CSI_CNN,
  author={Zhou, Rui and Hou, Huanhuan and Gong, Ziyuan and Chen, Zuona and Tang, Kai and Zhou, Bao},
  journal={IEEE Sens. J.}, 
  title={Adaptive Device-Free Localization in Dynamic Environments Through Adaptive Neural Networks}, 
  year={2021},
  volume={21},
  number={1},
  pages={548-559},
  doi={10.1109/JSEN.2020.3014641}}

@ARTICLE{CSI_attention,
  author={Shen, Li-Hsiang and Hsiao, An-Hung and Lu, Kuan-I and Feng, Kai-Ten},
  journal={IEEE Sens. J.}, 
  title={Attention-Enhanced Deep Learning for Device-Free Through-the-Wall Presence Detection Using Indoor {WiFi} Systems}, 
  year={2024},
  volume={24},
  number={4},
  pages={5288-5302},
  doi={10.1109/JSEN.2023.3346482}}

@ARTICLE{ML_RFSensing_Review,
  author={Santra, Avik and Wang, Pu and Shaker, George and Mysore, Bhavani Shankar and Dolmans, Guido and Chen, Yan and Shariati, Negin and Pandharipande, Ashish},
  journal={IEEE Sens. J.}, 
  title={Machine Learning-Powered Radio Frequency Sensing: A Review}, 
  year={2025},
  volume={25},
  number={13},
  pages={23164-23183},
  doi={10.1109/JSEN.2025.3547673}}

@ARTICLE{RSSI_meas,
  author={Xue, Weixing and Qiu, Weining and Hua, Xianghong and Yu, Kegen},
  journal={IEEE Sens. J.}, 
  title={Improved Wi-Fi RSSI Measurement for Indoor Localization}, 
  year={2017},
  volume={17},
  number={7},
  pages={2224-2230},
  doi={10.1109/JSEN.2017.2660522}}

@ARTICLE{RT_revirew,
  author={Fuschini, Franco and Vitucci, Enrico M. and Barbiroli, Marina and Falciasecca, Gabriele and Degli-Esposti, Vittorio},
  journal={Radio Sci.}, 
  title={Ray tracing propagation modeling for future small-cell and indoor applications: A review of current techniques}, 
  year={2015},
  volume={50},
  number={6},
  pages={469-485},
  doi={10.1002/2015RS005659}}

@ARTICLE{RT01,
  author={Tan, Jundong and Su, Zhuo and Long, Yunliang},
  journal={IEEE Trans. Antennas Propag.}, 
  title={A Full {3-D} {GPU}-based Beam-Tracing Method for Complex Indoor Environments Propagation Modeling}, 
  year={2015},
  volume={63},
  number={6},
  pages={2705-2718},
  doi={10.1109/TAP.2015.2415036}}

@ARTICLE{statistical_channel,
  author={Hashemi, H.},
  journal={Proc. IEEE}, 
  title={The indoor radio propagation channel}, 
  year={1993},
  volume={81},
  number={7},
  pages={943-968},
  doi={10.1109/5.231342}}

@ARTICLE{tomography,
  author={Nannuru, Santosh and Li, Yunpeng and Zeng, Yan and Coates, Mark and Yang, Bo},
  journal={IEEE Trans. Mob. Comput.}, 
  title={Radio-Frequency Tomography for Passive Indoor Multitarget Tracking}, 
  year={2013},
  volume={12},
  number={12},
  pages={2322-2333},
  doi={10.1109/TMC.2012.190}}

@ARTICLE{localization,
  author={Xue, Min and Sun, Wei and Yu, Hongshan and Tang, Hongwei and Lin, Anping and Zhang, Xing and Zimmermann, Roger},
  journal={IEEE Internet Things J.}, 
  title={Locate the Mobile Device by Enhancing the {WiFi}-Based Indoor Localization Model}, 
  year={2019},
  volume={6},
  number={5},
  pages={8792-8803},
  doi={10.1109/JIOT.2019.2923433}}

@ARTICLE{healthcare02,
  author={Chen, Jinbo and Zhang, Dongheng and Wu, Zhi and Zhou, Fang and Sun, Qibin and Chen, Yan},
  journal={IEEE Trans. Mob. Comput.}, 
  title={Contactless Electrocardiogram Monitoring With Millimeter Wave Radar}, 
  year={2024},
  volume={23},
  number={1},
  pages={270-285},
  doi={10.1109/TMC.2022.3214721}}

@techreport{Dogaru2007muscle,
  author      = {Dogaru, Traian and Nguyen, Lam H. and Le, Calvin},
  title       = {Computer Models of the Human Body Signature for Sensing Through the Wall Radar Applications},
  institution = {U.S. Army Research Laboratory},
  address     = {Adelphi, MD},
  number      = {ARL-TR-4290},
  year        = {2007},
  month       = sep,
  note        = {DTIC accession no. ADA473937},
}

@ARTICLE{2015TIM,
  author={Piuzzi, Emanuele and D’Atanasio, Paolo and Pisa, Stefano and Pittella, Erika and Zambotti, Alessandro},
  journal={IEEE Trans. Instrum. Meas.}, 
  title={Complex Radar Cross Section Measurements of the Human Body for Breath-Activity Monitoring Applications}, 
  year={2015},
  volume={64},
  number={8},
  pages={2247-2258},
  doi={10.1109/TIM.2015.2390811}}

\end{document}